\documentclass[final,1p,times]{elsarticle}

\usepackage{amsmath,amssymb,amsthm,mathtools}
\usepackage{bm}
\usepackage{graphicx}
\usepackage{placeins}
\usepackage{hyperref}
\usepackage{booktabs}
\usepackage{physics}
\usepackage{siunitx}
\usepackage{color}
\usepackage[utf8]{inputenc}
\usepackage[T1]{fontenc}
\usepackage[portuguese,english]{babel}
\usepackage{amssymb}
\usepackage{amsmath}
\usepackage{amsthm}
\usepackage{xcolor}
\usepackage{natbib}
\usepackage{hyperref}
\usepackage{subcaption}
\usepackage[utf8]{inputenc}
\usepackage[T1]{fontenc}
\usepackage{amssymb}
\usepackage{amsmath}
\usepackage{amsthm}
\usepackage{xcolor}
\usepackage{natbib}
\usepackage{hyperref}
\usepackage{subcaption}
\usepackage{booktabs}

\journal{Physics of the Dark Universe}

\begin{document}

\begin{frontmatter}



\title{Emergent inflation in fractional cosmology}

\author[1]{S. M. M. Rasouli}
\ead{mrasouli@ubi.pt}
 \address[1]{Departamento de F\'{i}sica,
Centro de Matem\'{a}tica e Aplica\c{c}\~{o}es (CMA-UBI),
Universidade da Beira Interior,
Rua Marqu\^{e}s d'Avila
e Bolama, 6200-001 Covilh\~{a}, Portugal}

\begin{abstract}
We investigate the dynamics of the early universe within an emergent fractional
cosmological framework generated by a measure-deformed Newtonian action,
without introducing additional propagating degrees of freedom.
Considering asymptotic and dynamical requirements, we construct a minimal
fractional potential that, together with the time-dependent fractional
kernel, generates a regular pre-inflationary stage, quasi-de Sitter
inflation, a dynamically controlled exit, and a subsequent finite duration radiation-like phase. The exact solution of the reduced early-time
equation is obtained in two equivalent representations, and regularity
selects a unique physical branch with finite principal curvature invariants
at the initial boundary. Remarkably, the same exact trajectory evolves
asymptotically toward the quasi-de Sitter regime, while an exact
nonlinear analysis of homogeneous Hubble deviations establishes its
asymptotic attractor character within the admissible expanding sector.
Inflation ends dynamically near the zero crossing of the effective
fractional force, leading to controlled background-level consistency
relations among the inflationary duration, the fractional parameter,
the effective equation of state parameter, and the exit scale. For the physically
relevant post-inflationary regime, the model admits the
characteristic radiation-like expansion over a finite post-inflationary interval.
These results show that a coherent sequence of regular pre-inflationary,
inflationary, and post-inflationary background dynamics can emerge from
a simple and unified fractional Newtonian framework without invoking an
inflaton field.
\end{abstract}


\end{frontmatter}

\section{Introduction}
\label{SecI}

Various inflationary models have been proposed as frameworks for describing the early universe~\cite{guth1981inflationary,guth1984inflationary,linde2000inflationary}. Specific realizations include the Starobinsky model~\cite{vilenkin1985classical}, hybrid inflation~\cite{linde1994hybrid}, and k-inflation~\cite{armendariz1999k}. Other developments address eternal inflation~\cite{guth2000inflation}, natural inflation~\cite{freese2004natural}, and chaotic inflation in supergravity~\cite{kallosh2010new}. Inflation has also been investigated within fractional quantum cosmology~\cite{rasouli2022inflation}. A comprehensive review of inflationary models is given in Ref.~\cite{martin2024encyclopaedia}.

Inflationary models provide a well-established framework for addressing fundamental problems of standard cosmology, such as the horizon, flatness, and 
large-scale homogeneity problems~\cite{guth1981inflationary,linde1982new}. Further developments include eternal extended inflation with a graceful exit~\cite{linde1990eternal} and Higgs inflation, in which the Standard Model Higgs field acts as the inflaton through a nonminimal coupling to gravity~\cite{bezrukov2008standard}.
 Related investigations have considered horizon-problem remediation through deformed phase space~\cite{rasouli2011horizon} and the role of the Jordan and Einstein frames in inflationary magnetogenesis~\cite{velasquez2024jordan}.

Moreover, inflation offers a natural mechanism for generating primordial fluctuations that seed the observed large-scale structure of the universe, as discussed in general treatments of inflationary perturbations~\cite{mukhanov2007cosmological,baumann2009tasi,langlois2010inflation}. 
Specific studies have examined the evolution of gravitational perturbations in non-commutative inflation~\cite{koh2007evolution} and the effects of nonminimal coupling in chaotic inflation~\cite{fakir1990improvement}.
Despite these phenomenological successes, the physical origin of inflation and the fundamental mechanism driving this phase remain only partially understood, with open issues discussed in Refs.~\cite{martin2024encyclopaedia,Brandenberger:2001zqv,ijjas2013inflationary}. These questions motivate the exploration of alternative and complementary theoretical frameworks. Examples include gravity-driven acceleration in a noncommutative Brans–Dicke setting~\cite{rasouli2016gravity}, power-law inflation associated with phase-space noncommutativity~\cite{rasouli2024phase}, and noncommutative constant-roll inflation with nonminimal coupling~\cite{gashti2025noncommutativity}.

A large class of inflationary models relies on the introduction of an
effective scalar field, commonly referred to as the inflaton, together
with a suitably chosen potential governing its dynamics
\cite{linde1983chaotic,linde2000inflationary,brandenberger2010cosmology}.
Although this approach has proven to be phenomenologically successful,
the strong dependence of inflationary dynamics on the assumed form of
the potential, along with the absence of direct experimental evidence
for the inflaton as a fundamental scalar degree of freedom, raises
important conceptual questions regarding the physical nature of
inflation and the true identity of its underlying degrees of freedom
\cite{martin2024encyclopaedia,baumann2009tasi,borde2003inflationary,meerburg2020planck}.
Moreover, the wide variety of proposed scalar potentials and the
sensitivity of inflationary predictions to their specific forms suggest
that a scalar-field description may not represent the unique or most
fundamental realization of inflationary dynamics.
These considerations motivate the exploration of alternative formulations of early-universe dynamics. Examples include the Starobinsky model and its classical and quantum descriptions~\cite{vilenkin1985classical,starobinsky1980new}, nonlocal cosmology~\cite{deser2007nonlocal}, and constant-roll inflation in (f(R)) gravity~\cite{nojiri2017constant}. Other approaches investigate multifractional spacetimes~\cite{calcagni2012introduction} and gravitational theories with fractional operators~\cite{calcagni2021classical}. The effective field theory of inflation provides a complementary framework for describing inflationary dynamics~\cite{cheung2008effective}.

In this context, a variety of non-standard approaches have been explored in recent years. Fractional cosmological models have been studied through modifications of the Einstein–Hilbert action~\cite{shchigolev2013fractional} and through exact solutions and cosmological constraints~\cite{gonzalez2023exact}. Quantum cosmological applications include fractional formulations~\cite{rasouli2021broadening}, fractional k-essence~\cite{socorro2023quantum}, and quantum creation of an FLRW universe using the Riesz fractional derivative~\cite{canedo2025quantum}. Other studies derive modified Friedmann equations from fractional entropy~\cite{ccoker2023modified} and a fractal apparent horizon~\cite{jalalzadeh2024friedmann}, or constrain fractal Chern–Simons gravity using astronomical observations~\cite{el2024constraining}. Related applications in fractional quantum mechanics examine tunnelling through Cantor potentials~\cite{singh2023quantum}.
Among these, fractional frameworks have attracted increasing attention
due to their intrinsic ability to incorporate temporal non-locality and
memory effects
\cite{garcia2022cosmology,micolta2023revisiting,benetti2023dark,contreras2025fractional}.
Fractional formulations have been applied across different areas of
physics, ranging from classical and quantum mechanics to gravity and
cosmology
\cite{gonzalez2023exact,costa2023estimated,rasouli2025gravitational,micolta2025fractional}.
In cosmological settings, fractional modifications have been shown to
affect the description of singularities, cosmic phase transitions, and
effective energy-momentum behavior across different epochs of the
universe, from early-time dynamics to late-time accelerated expansion
\cite{rasouli2022inflation,costa2023estimated, shchigolev2021fractional,rasouli2024fractional}.

Within this perspective, the main goal of this work is to investigate
several fundamental questions that naturally arise as follows.
Can the \textit{emergent $\Lambda$CDM cosmology\footnote{We also refer
to this framework as ``\textit{fractional Newtonian cosmology (NC)}''.}}
\cite{rasouli2026emergent} (which can be viewed as an effective reduced
realization of fractal cosmological frameworks with generalized
measures, where the standard integral measure is replaced by a
Stieltjes-type measure
\cite{calcagni2010fractal,calcagni2010quantum}),
provide a self-consistent and structured description of the very early
universe?
Is it possible, within such a framework, to reproduce key background
features associated with conventional inflationary models while
avoiding the introduction of fundamental scalar fields?
Can fractional modifications of the dynamics give rise to a regular
pre-inflationary phase and dynamically generate a quasi-de Sitter
inflationary regime that acts as an asymptotic background attractor?
Furthermore, can such a scenario naturally admit a dynamically
controlled exit from inflation followed by a finite-duration radiation-like phase, while generating a substantial number
of $e$-folds?
The following sections investigate these questions within the fractional
cosmological framework established in
Ref.~\cite{rasouli2026emergent}. Rather than representing an isolated
inflationary construction, the present work constitutes the
early-universe sector of the broader fractional cosmological program.
Together with the radiation-dominated, matter-dominated, and late-time
accelerating solutions obtained in
Ref.~\cite{rasouli2026emergent}, the present results provide a
background-level description spanning the early-universe dynamics
investigated here and the subsequent cosmological regimes within the
same theoretical framework.

The paper is organized as follows. 
In the next section, we review the basic framework of fractional NC
(emergent $\Lambda$CDM cosmology) and derive the corresponding cosmological equations from the fractional action.
In Section~\ref{Inf-assump}, we construct a minimal fractional potential suitable for
describing the early-universe dynamics.

In Section~\ref{inflation}, we investigate the emergence of the quasi-de Sitter
inflationary regime.
In Section~\ref{pre-inf}, we obtain the exact solution of the reduced background equation in two equivalent
representations. After selecting the regular branch, we investigate its early-time
curvature regularity, and establish its asymptotic connection to the
quasi-de Sitter inflationary regime.
In Section~\ref{Dyn-pre-infl}, we investigate the dynamics of the
pre-inflationary--inflationary transition and analyze the asymptotic
attractor character of the inflationary background under homogeneous
Hubble deviations.
Section~\ref{Post-inf} is devoted to the post-inflationary regime, including a finite-duration radiation-like background, the dynamically controlled exit from
inflation, and the complete-background and finite-time contractivity
analysis of the radiation-like entrance. \Ref{AppendixA} clarifies the
energy normalization and matching of the fractional contributions
across the regular pre-inflationary, inflationary, and radiation-like
regimes, while \Ref{AppendixB} summarizes the time, endpoint, and e-fold
conventions used throughout the analysis.
Finally, Section~\ref{concl} summarizes the main results and presents a comparative
discussion with standard scalar-field, Starobinsky
\cite{vilenkin1985classical,starobinsky1980new}, and $\alpha$-attractor
inflationary scenarios \cite{kallosh2013universality}.


\section{Measure-induced fractional dynamics and the emergent cosmological equations}
\label{FNC}

In this section, we briefly review the emergent $\Lambda$CDM cosmology arising from a measure-induced deformation of the Newtonian action, also referred to as fractional Newtonian cosmology (NC) \cite{rasouli2026emergent}. 

 As we discussed in \cite{rasouli2026emergent}, the introduction of a time-dependent kernel effectively replaces the standard integration measure with a generalized Stieltjes-type measure, allowing memory effects and history dependence to be incorporated directly into the dynamics. From a conceptual standpoint, this construction is aligned with broader approaches in which non-trivial measures are employed to represent generalized geometric or scale-dependent structures \cite{calcagni2010fractal,calcagni2010quantum}. 
However, in \cite{rasouli2026emergent}, the fractional  modification serves as minimal phenomenological implementation of temporal weighting in the gravitational sector. In this sense, the friction-like term appearing in the equations of motion can be understood as an effective manifestation of memory effects induced by the underlying measure structure, without introducing additional propagating degrees of freedom. 
It is worth emphasizing that this minimal modification introduces only
a single constant fractional parameter $\alpha$, yet the resulting
framework is capable of reproducing key dynamical features of the universe at both the background and perturbative levels \cite{rasouli2026emergent,rasouli2026evolution}, comparable to effective relativistic models, while the standard Newtonian description remains essentially restricted to the matter-dominated regime and fails to capture other cosmological phases.

In Ref.~\cite{rasouli2026emergent}, the classical (Newtonian) action was modified as
\begin{eqnarray}\label{New-fr-action}
 S_{\alpha}=
 \!\!\frac{1}{\Gamma(\alpha)}\int_{0}^{\bar t} \xi({\tau} )
 \left(T-V_{\rm eff}\right)\,d{\tau},
\end{eqnarray}
with
\begin{equation}\label{L-xi}
 \xi({\tau})\equiv\left(\frac{{\bar t} - \tau}{\tau_0}
\right)^{\alpha-1},
\end{equation}
where $\tau_0$ denotes a fixed reference time scale to solely make the time-dependent kernel $ \xi$ dimensionless.
Within this framework, $T=\frac{1}{2}\,m (\frac{d {\mathbf{r}}}{d\tau} )^2$ is the standard kinetic energy, and $V_{\rm eff}=V_{\rm eff}(\mathbf{r};\alpha)$ denotes the effective potential energy of the system. 
A central assumption of the model is that the fractional parameter
$\alpha$ is a positive real number. In the special case $\alpha=1$,
the effective potential reduces to the standard Newtonian potential
$V(r)$ associated with conservative forces.

The equations of motion corresponding to the action \eqref{New-fr-action} are given by 
\begin{eqnarray}\label{fr-eq-3d}
m\,\ddot{\mathbf{r}}
= -\nabla V_{\rm eff}(\mathbf{r})- m \gamma_\alpha(t) \,\dot{\mathbf{r}}, \hspace{5mm} \gamma_\alpha(t)\equiv \frac{\dot \xi}{\xi}=  \frac{(\alpha - 1)}{t},
\end{eqnarray}
where we set $\bar{t}-\tau\equiv t$ and $\nabla$ is the gradient operator with respect to the particle position $\mathbf r$. 
Moreover, $t$ denotes the cosmic time employed in the dynamical and
cosmological equations, and an overdot denotes differentiation with
respect to $t$.

Taking the scalar product of equation~\eqref{fr-eq-3d} with
$\dot{\boldsymbol{r}}$ gives
\begin{equation}\label{proof1}
\frac{d}{dt}\left(T+V_{\rm eff}\right)
=
-2\gamma_{\alpha}(t)T,
\end{equation}
where $T=m\dot{\boldsymbol{r}}^{\,2}/2$. We therefore introduce
the fractional kinetic contribution through the differential
relation 
\begin{equation}
\dot{T}_{\alpha}(t)
\equiv
2\gamma_{\alpha}(t)T(t).
\label{Talpha-differential}
\end{equation}
With this definition,
\begin{eqnarray}\label{Fr-cons}
\mathcal{E^{^{\rm mech}}_{\rm eff}}\equiv T + V_{\rm eff} +T_\alpha = \text{constant}.
\end{eqnarray}

The general integrated
form of equation~\eqref{Talpha-differential} is
\begin{equation}\label{frac-T}
T_{\alpha}(t)
=
C_T(\alpha)
+
2(\alpha-1)
\int_{t_0}^{t}
T(t')\,d\ln t'
\end{equation}
where $t_0$ is a reference time and $C_T(\alpha)$ is an
additive integration constant, whose value is fixed by the adopted energy
convention once the relevant initial or matching conditions are
specified. In the special Newtonian case, $\alpha=1$, we
set $C_T(1)=0$, so that the fractional kinetic contribution
vanishes.

Within the above fractional dynamics, the relativistic 
cosmological equations are shown to emerge effectively.

We consider a spherical region of  physical radius $r$, filled with pressure-less matter of uniform density $\rho_{\rm m}$. A test particle of mass $m$ is located on the surface of the sphere and is subject to the Newtonian
gravitational potential $\Phi_N$, which obeys
\begin{equation}\label{N-Pot}
    \nabla_{\!r}\,\Phi_{\rm N}=\frac{4 \pi G}{3} \rho_m \mathbf{r}.
\end{equation}

From this setup, the standard Newtonian cosmological equations can be derived, following the foundational treatments of Milne and McCrea~\cite{milne1934newtonian,mccrea1934newtonian,mccrea1955newtonian,mccrea1955significance}. Further studies discuss cosmological dynamics within Newtonian mechanics~\cite{callan1965cosmology,jordan2005cosmology}, discrete Newtonian cosmology~\cite{ellis2013discrete}, and a re-examination of Newtonian cosmology~\cite{tipler1996newtonian}. The relation between the Newtonian analogy and relativistic Friedmann--Lema\^{i}tre--Robertson--Walker (FLRW) cosmology was investigated in Ref.~\cite{Faraoni:2020uuf}.

However, in our fractional cosmological framework, a modified version of the above construction was established~\cite{rasouli2026emergent}.

In the cosmological specialization, we write
$r(t)=a(t)\varrho$, where $a(t)$ is the scale factor and $\varrho$ is the
constant comoving radial coordinate of the test particle. For a fixed
comoving shell, the radial coordinate can be normalized to $\varrho=1$,
so that $r(t)=a(t)$ and $\dot r=\dot a$. We adopt this normalization
in what follows.

In what follows, we summarize and re-derive
the main cosmological equations of the measure-induced fractional
Newtonian framework.

It is convenient to define
\begin{eqnarray}\label{cos-E-def}
E&\equiv& T+\Phi_{_{N}}, \hspace{5mm} E_{\alpha}\equiv T_{\alpha}+\Phi_{\alpha},
\end{eqnarray}
where $T\equiv1/2\, \dot a^2$, and $ T_{\alpha}$ denotes the fractional kinetic contribution defined by equation \eqref{frac-T}. In the cosmological specialization considered here, we decompose the effective gravitational potential as
\begin{eqnarray}\label{eff-phi}
\Phi_{\rm eff}(a)\equiv V_{\rm eff}/m=\Phi_{_{N}}+\Phi_{\alpha}.
\end{eqnarray} Accordingly, the conserved effective 
mechanical energy introduced in equation \eqref{Fr-cons} takes the form
\begin{equation}\label{Etotal}
\mathcal E
\equiv
E+E_{\alpha}
=
{\rm constant}.
\end{equation}

Using equations \eqref{N-Pot}-\eqref{Etotal}, the conserved energy relation can be written as
\begin{equation}\label{energy_cosmology}
\frac{1}{2}\dot a^{\,2}
-
\frac{4\pi G}{3}\rho_m a^2
+
E_{\alpha}
=
\mathcal E.
\end{equation}

Multiplying both sides of 
equation \eqref{energy_cosmology} by $2/a^2$ and defining $H\equiv \dot a/a$, we obtain
\begin{equation}\label{H_before_def}
H^2
=
\frac{8\pi G}{3}\rho_m
-
\frac{2E_{\alpha}}{a^2}
+
\frac{2\mathcal E}{a^2}.
\end{equation}

We now define
\begin{equation}\label{rho-eff}
\mathcal{K}\equiv-2\mathcal E,\qquad \rho_{\alpha}
\equiv
-\frac{3}{4\pi G}
\frac{E_{\alpha}}{a^2},
\end{equation}
where $\rho_{\alpha}$ depends solely on the fractional energy 
$E_{\alpha}$ and is not an independent fluid component. 

Equation~\eqref{H_before_def} therefore takes the Friedmann-like form
\begin{equation}\label{eff-frd}
H^2
=
\frac{8\pi G}{3}\rho_{\rm eff}
-
\frac{\mathcal{K}}{a^2},
\qquad
\rho_{\rm eff}
\equiv
\rho_m+\rho_{\alpha}.
\end{equation}

In what follows, we derive the fractional acceleration equation. Using \eqref{frac-T}, one can easily show that $\dot T_{\alpha}=\gamma_{\alpha}\dot a^{\,2}$, 
and therefore, along the cosmological trajectory, ${dT_{\alpha}}/{da}
={\dot T_{\alpha}}/{\dot a}=\gamma_{\alpha}\dot a.$
Consequently,
\begin{equation}\label{dEalpha_da}
\frac{dE_{\alpha}}{da}
=
\gamma_{\alpha}\dot a
+
\frac{d\Phi_{\alpha}}{da}.
\end{equation}

The cosmological version of the fractional equation of motion \eqref{fr-eq-3d} is
\begin{equation}\label{cosmological_EOM}
\ddot a
+
\gamma_{\alpha}\dot a
+
\frac{d\Phi_N}{da}
+
\frac{d\Phi_{\alpha}}{da}
=0.
\end{equation}
Using equations \eqref{N-Pot} and \eqref{dEalpha_da}, equation \eqref{cosmological_EOM} becomes
\begin{equation}\label{acceleration_Ealpha}
\frac{\ddot a}{a}
=
-\frac{4\pi G}{3}\rho_m
-
\frac{1}{a}\frac{dE_{\alpha}}{da}.
\end{equation}

To express the fractional contribution in an effective fluid form, we
define $p_\alpha$ as the unique effective pressure for which the
density $\rho_\alpha$ introduced in equation~\eqref{rho-eff} satisfies the
standard continuity equation along the cosmological trajectory:
\begin{equation}\label{p-eff}
p_{\alpha}
\equiv
\frac{1}{4\pi G a^2}
\frac{d}{da}
\left(
aE_{\alpha}
\right),
\end{equation}
Equations \eqref{rho-eff} and \eqref{p-eff} show that, once $E_{\alpha}$ is specified, both
$\rho_{\alpha}$ and $p_{\alpha}$ are completely determined and are not independent fluid quantities.

Indeed, combining equations~\eqref{rho-eff} and~\eqref{p-eff} gives
\begin{align}\label{rho3p_identity}
\rho_{\alpha}+3p_{\alpha}
=
-\frac{3E_{\alpha}}{4\pi G a^2}
+
\frac{3}{4\pi G a^2}
\frac{d}{da}(aE_{\alpha})
=
\frac{3}{4\pi G a}
\frac{dE_{\alpha}}{da}.
\end{align}
Hence
\begin{equation}\label{Ealpha_rhop}
\frac{1}{a}\frac{dE_{\alpha}}{da}
=
\frac{4\pi G}{3}
\left(
\rho_{\alpha}+3p_{\alpha}
\right).
\end{equation}

Substituting \eqref{Ealpha_rhop} into equation~\eqref{acceleration_Ealpha} yields
\begin{equation}\label{acc-eff-eq}
\frac{\ddot a}{a}
=
-\frac{4\pi G}{3}
\left(
\rho_{\rm eff}+3p_{\rm eff}
\right),
\qquad
p_{\rm eff}\equiv p_{\alpha}.
\end{equation}

For completeness, the conservation equations can also be derived
directly. 

(i) Combining equations \eqref{eff-frd} and \eqref{acc-eff-eq} yields a continuity equation for the effective sector as
\begin{eqnarray}\label{cons-eq-cos}
\dot{\rho}_{_{\rm eff}}+3H (\rho_{_{\rm eff}}+p_{_{\rm eff}})=0.
\end{eqnarray}

(ii) Independently, differentiating both sides of  \eqref{rho-eff} with respect to the cosmic time gives
\begin{align}\label{rhoa_dot_1}
\dot\rho_{\alpha}
=
-\frac{3}{4\pi G}
\frac{d}{dt}
\left(
\frac{E_{\alpha}}{a^2}
\right)
=
-\frac{3}{4\pi G}
\left[
\frac{\dot E_{\alpha}}{a^2}
-
\frac{2E_{\alpha}\dot a}{a^3}
\right],
\end{align}
which can be rewritten as
\begin{equation}\label{rhoa_dot_2}
\dot\rho_{\alpha}
=
-\frac{3H}{4\pi G a^2}
\left(
a\frac{dE_{\alpha}}{da}
-
2E_{\alpha}
\right).
\end{equation}
 On the other hand, equations \eqref{rho-eff} and \eqref{p-eff} imply
\begin{equation}\label{rhoa_plus_pa}
\rho_{\alpha}+p_{\alpha}
=
\frac{1}{4\pi G a^2}
\left(
a\frac{dE_{\alpha}}{da}
-
2E_{\alpha}
\right).
\end{equation}

Combining equations~\eqref{rhoa_dot_2} and~\ref{rhoa_plus_pa}, one obtains identically
\begin{equation}\label{cons-eq-fr}
\dot\rho_{\alpha}
+
3H
\left(
\rho_{\alpha}+p_{\alpha}
\right)
=0.
\end{equation}

(iii) Finally, using \eqref{eff-frd} and the 
conservation equations \eqref{cons-eq-cos} and \eqref{cons-eq-fr}, we 
obtain the conservation equation for the ordinary matter as 
\begin{eqnarray}\label{cons-eq-rho-m}
\dot{\rho}_{\rm m}+3H\rho_{\rm m}=0,
\end{eqnarray}
which is the same equation valid in the standard NC.

In summary, equations~\eqref{eff-frd}, \eqref{acc-eff-eq} and \eqref{cons-eq-cos} are the main dynamical equations of motion associated with the fractional
Newtonian cosmology. In this fractional model, $\rho_\alpha$ is
introduced from the fractional energy through equation~\eqref{rho-eff}, while
$p_\alpha$ is defined by equation~\eqref{p-eff} as the unique effective
pressure satisfying the continuity equation~\eqref{cons-eq-fr}. With these
definitions, the acceleration equation~\eqref{acc-eff-eq} and the effective
continuity equation~\eqref{cons-eq-cos} follow algebraically. No additional
independently propagating fluid degree of freedom is thereby
introduced.


Before proceeding, we highlight several key features of the fractional
cosmological framework established in Ref.~\cite{rasouli2026emergent}, which are particularly relevant to the present analysis.

\begin{itemize}

\item
The cosmological evolution is not determined solely by the effective potential. The effective energy density and pressure depend on the total fractional contribution $E_{\alpha}=T_{\alpha}+\Phi_{\alpha}$, where $T_{\alpha}$ originates from the time-dependent kernel $\xi$ and $\Phi_{\alpha}$ represents the fractional potential. Consequently, the relative importance of these two contributions changes throughout cosmic evolution.

\item

The ordinary matter component remains a pressure-less dust fluid, as in
standard Newtonian cosmology, satisfying $p_m=0$ and $\dot{\rho}_m+3H\rho_m=0$. 
This choice is not an additional restriction of the present model, but rather reflects the structure of the Newtonian cosmological construction, in which the matter source entering the Poisson equation is a non-relativistic mass density. 
However, the total effective cosmological dynamics is not dust dominated in general. 
The fractional sector contributes an induced energy density $\rho_\alpha$ and an effective pressure $p_\alpha$, so that $p_{\rm eff}=p_\alpha$ may differ from zero even though the ordinary matter pressure vanishes. 
Consequently, radiation-like expansion, inflationary acceleration, and late-time accelerated expansion (which cannot be described within the standard NC) arise from the measure-induced fractional modification of the gravitational dynamics.
The pressureless matter contribution is present throughout the entire cosmological evolution. However, as in the standard relativistic cosmological description, different epochs are characterized by different dominant contributions to the expansion dynamics. 

\item
The fractional NC formalism and the exact background solutions corresponding to the radiation-dominated, matter-dominated, and late-time accelerating phases have been comprehensively presented in Ref.~\cite{rasouli2026emergent}. The corresponding perturbations have been systematically analyzed  in Ref.~\cite{rasouli2026evolution}.
In the following sections, we extend the cosmological applications of the emergent cosmological framework by analyzing the dynamics of the early universe.

\end{itemize}


\section{Physical requirements and construction of the minimal fractional potential for early-universe evolution}
\label{Inf-assump}

In what follows, an appropriate potential for describing the evolution of the early universe is introduced, and in the subsequent sections, three successive phases and the dynamical transitions between them are investigated within the framework of fractional cosmology.
The aim of the present paper is not to reconstruct the full sequence of late-time cosmological epochs once again, since the radiation-dominated, matter-dominated, and late-time accelerating background solutions have already been obtained within the emergent $\Lambda$CDM framework of Ref.~\cite{rasouli2026emergent}. 
Here, we instead apply the same fractional Newtonian framework to investigate the evolution of the very early universe.

Let us assume $\mathcal{K}=0$, which corresponds to the spatially flat FLRW universe and is in agreement with recent observational data \cite{Planck:2018vyg}.
This spatially flat FLRW condition is implemented through the conserved
energy normalization and does not require the additive constants in
$T_\alpha$ and $\Phi_\alpha$ to vanish separately. Their consistent
normalization along the regular, inflationary, and radiation-like
representations is derived in \ref{AppendixA}.

As mentioned in the previous section, throughout this work we assume that the effective potential can be decomposed into Newtonian and fractional contributions, such that $\Phi_{\alpha}=0$ in the special case $\alpha=1$. 

For later convenience, instead of using the fractional acceleration equation \eqref{acc-eff-eq}, let us use its equivalent form, equation \eqref{cosmological_EOM}. Substituting $\Phi_{\rm N}$ from \eqref{N-Pot} into equation \eqref{cosmological_EOM}, we obtain 
\begin{eqnarray}\label{acc-eq}
\frac{\ddot a}{a} =\dot{H}+H^2=-\gamma_\alpha(t)\,H-\frac{4\pi G}{3} \rho_{\rm m}(a)+\mathcal{F}(a),
\end{eqnarray}
where we defined $\mathcal{F}(a)$ as
 \begin{eqnarray}\label{K-phi}
\mathcal{F}(a) \equiv -\frac{1}{a}\frac{\partial\Phi_{\alpha}}{\partial a}.
\end{eqnarray}

Equation \eqref{acc-eq} also clarifies the role of the dust component in the present construction. 
The term proportional to $\rho_m(a)$ is the ordinary  matter contribution and, by virtue of equation~\eqref{cons-eq-rho-m}, always scales as $\rho_m\propto a^{-3}$. 
It is therefore responsible for the standard matter-dominated regime already recovered in the full emergent $\Lambda$CDM construction.

In this work, we propose a fractional scenario that describes the dynamics of the early universe. 
Equation \eqref{acc-eq} contains the time-dependent fractional term
$-\gamma_\alpha(t)H$. We will show that although $\gamma_\alpha(t)$ diverges as
$t\to0^+$, the regular solution satisfies $H\propto t$, so the product
$\gamma_\alpha H$ remains finite. The early-time dynamics should
therefore be inferred from the regular solution of the full reduced
equation, rather than described as being dominated by a divergent
fractional term. Accordingly, the effective force $\mathcal F(a)$
should be chosen such that it drives the inflationary phase and behaves
appropriately in the neighbouring regimes.

Let us be more precise. In order to propose a suitable form of $\mathcal{F}(a)$ that satisfies the criteria mentioned above, we assume a constant characteristic scale factor $a_c>0$, whose magnitude is assumed to be 
$ a_c \sim a_{\rm end}$, where $a_{\rm end}$ denotes the magnitude of the scale factor at the end of inflation. The function $\mathcal{F}(a)$ should then satisfy the following requirements:
\begin{itemize}
    \item 
Suppose that we can write $\mathcal{F}$ as a function of $x$, where $x=x(a/a_{\rm c})$. For $x(a/a_{\rm c})\ll1$, the force should approach a positive constant, $\mathcal{F}\to \mathcal{F}_0>0$, generating an approximately de Sitter expansion at early times. 

\item
For $x\gg1$, it should reduce as a power law $\mathcal{F}(a)\propto -\mathcal{F}_0/a^4$, which supports a radiation-like post-inflationary background. 
\end{itemize}

In this respect, let us propose the fractional potential as

\begin{equation}\label{phi-alpha-def}
\Phi_{\alpha}(a)
=
-\Phi_0(\alpha) \left[\frac{a^2}{1+(\frac{a}{a_c})^p}\right]
+
C_{\Phi}(\alpha),
\end{equation}
where $C_{\Phi}(\alpha)$ denotes an $\alpha$-dependent constant. To recover the
standard Newtonian model in the special case $\alpha=1$, we
impose $\Phi_{0}(1)=0=C_{\Phi}(1)$ so that $\Phi_{\alpha}(a)|_{\alpha=1}=0$. Since
$C_{\Phi}(\alpha)$ is independent of the scale factor, it does
not contribute to the effective force
$\mathcal{F}(a)=-(1/a)\,d\Phi_{\alpha}/da$. It therefore leaves
the acceleration equation and the force-based dynamical
analysis unchanged. Nevertheless, it contributes to the total
fractional energy $E_{\alpha}=T_{\alpha}+\Phi_{\alpha}$ and
must consequently be fixed jointly with the additive constant
$C_T(\alpha)$ by the adopted energy convention and the initial
or matching conditions.

Moreover, $\Phi_0(\alpha)$ represents an effective dynamical amplitude 
characterizing the strength of the fractional gravitational sector. 
Dimensionally, it has units of inverse time squared, i.e. 
$[\Phi_0] = T^{-2}$, analogous to $H^2$ in standard cosmology. 

Substituting $\Phi_{\alpha}$ into \eqref{K-phi}, we obtain the corresponding 
interpolating function:
\begin{eqnarray}\label{K-def}
\mathcal{F}(x)=\mathcal{F}_0\left[\frac{1}{1+x}-\frac{p \,x}{2(1+x)^2}\right], \hspace{5mm}
x\equiv \left(\frac{a}{a_c}\right)^p, \hspace{5mm} \mathcal{F}_0(\alpha)\equiv2 \,\Phi_0(\alpha).
\end{eqnarray}

\begin{figure}
\centering\includegraphics[width=2.5in]{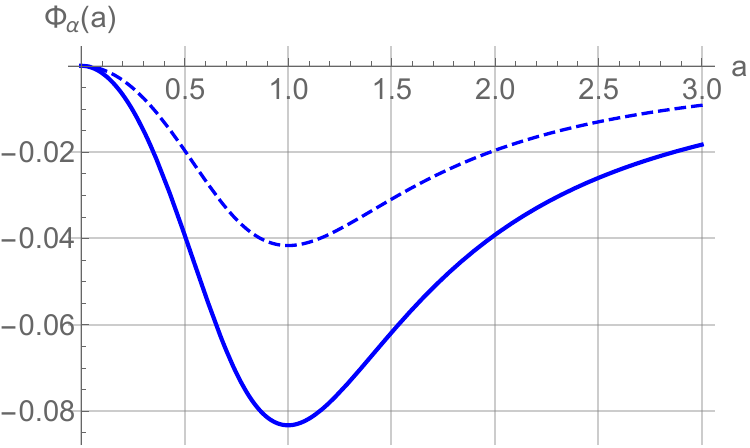}
\centering\includegraphics[width=2.5in]{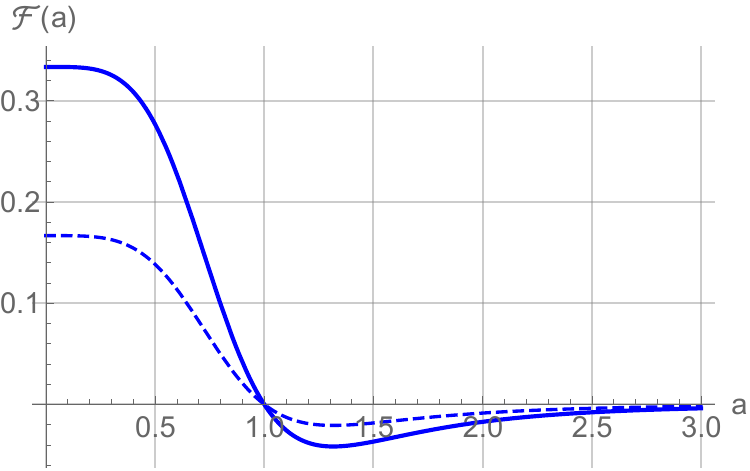}
\caption{The $\alpha$-dependent potential $\Phi_{\alpha}(a)$ (left panel) and the corresponding effective force $\mathcal{F}(a)$ (right panel) as functions of the scale factor. The solid and the dashed curves correspond to $\Phi_0=1/6$ and $\Phi_0=1/12$, respectively, and we have set $p= 4$, $a_c=1$ and $C_{\Phi}=0$.
}
\label{pot-F}
\end{figure}

As an illustration, in Figure~\ref{pot-F} we depict the
behavior of $\Phi_\alpha(a)$ and $\mathcal{F}(a)$ as functions
of the scale factor, adopting the illustrative normalization
$C_\Phi(\alpha)=0$. Any other choice of $C_\Phi(\alpha)$ would
only shift the potential vertically and would not affect the
corresponding force or the subsequent force-based dynamical
analysis.

At this stage, it is important to clarify the physical status of the fractional potential introduced above. 
The conditions $\Phi_0(1)=0$ and $C_\Phi(1)=0$ specify the
special Newtonian case exactly\footnote{In the general statements made in
Refs.~\cite{rasouli2026emergent,rasouli2026evolution}, the expression ``the Newtonian
limit $\alpha\to1$'' was intended to denote the exact special case
$\alpha=1$; accordingly, the vanishing of the fractional potentials
in those statements refers to this exact specialization.}
. No limiting behavior of
$\Phi_0(\alpha)$, $\mathcal{F}_0(\alpha)$, or
$C_\Phi(\alpha)$ as $\alpha\to1$ is assumed. In particular,
$\mathcal{F}_0(\alpha)\to0$ does not follow solely from
$\mathcal{F}_0(1)=0$.

The potential \eqref{phi-alpha-def} should not be interpreted as a fundamental interaction derived from an underlying microscopic theory.
Rather, it plays the same effective role that inflaton potentials play in conventional inflationary cosmology.
Indeed, in many conventional inflationary scenarios, the scalar
potential is introduced phenomenologically. Its functional form is
selected according to the cosmological dynamics it is intended to
reproduce, rather than being uniquely derived from first principles.
The philosophy adopted in the present work is conceptually analogous, although the underlying physical framework is fundamentally different.
Instead of introducing a scalar degree of freedom, the inflationary dynamics emerges from the fractional gravitational sector induced by the generalized measure.
Consequently, the fractional potential should be regarded as an effective description of this sector rather than as a fundamental interaction.

More importantly, the potential is not introduced as an arbitrary deformation of Newtonian gravity.
Instead, we first impose a set of physical and mathematical consistency requirements on the fractional cosmological dynamics and subsequently seek the simplest effective realization satisfying all of them.
In particular, the potential is required to (i) recover the standard Newtonian model in the exact special case
$\alpha=1$, for which $\Phi_\alpha|_{\alpha=1}=0$; (ii) as will be demonstrated in the following sections, generating a quasi-de Sitter expansion at early times; (iii) naturally provide a graceful exit from inflation; (iv) reproduce a radiation-like post-inflationary background; and (v) remain compatible with the observational constraints on the fractional parameter obtained within the same theoretical framework.
These requirements considerably restrict the admissible functional behaviour of the effective force and therefore substantially reduce the freedom in constructing the corresponding potential.
Consequently, the potential adopted in this work should be viewed as a minimal representative of a broader class of effective fractional potentials sharing the same asymptotic physical properties.
This is conceptually analogous to attractor inflationary models, where an entire class of distinct potentials may lead to essentially the same inflationary background evolution despite their different microscopic realizations; see, e.g., Ref.~\cite{galante2015unity}.

Finally, we emphasize that the characterization of the adopted potential
as a \textit{minimal representation} is intended in a phenomenological rather
than a strict mathematical sense. The proposed form provides a simple
realization of the asymptotic and dynamical requirements specified above
and is not claimed to constitute a unique interpolating potential.
The construction is motivated by our previous analysis of the same
fractional Newtonian framework, in which the effective fractional
potentials associated with the radiation-dominated, matter-dominated,
and late-time accelerating regimes were reconstructed explicitly from
the corresponding background solutions~\cite{rasouli2026emergent}. The
present potential is inspired by these reconstructed structures and
extends the same phenomenological strategy to the early-universe
dynamics considered here. Other smooth interpolating potentials
preserving the same essential asymptotic properties may therefore be
considered within the same framework. In particular, sufficiently small
smooth deformations that preserve the positive early-time regime, the
transition structure, and the required large-$a$ asymptotic behavior
are expected to retain the same qualitative sequence of inflation,
exit, and radiation-like post-inflationary expansion. The precise
transition scale and other quantitative details may, however, depend on
the particular interpolation adopted. A systematic classification of
such deformations lies beyond the scope of the present work.

In the following sections, we show that the potential \eqref{phi-alpha-def} satisfies the desired conditions.
  

\section{Emergence of the inflationary phase}
\label{inflation}

Before proceeding with the explicit solutions, it is useful to emphasize
the main objective and conceptual significance of the present analysis.
The inflationary scenario developed in this work is not intended merely
as an alternative phenomenological realization of inflation. Rather,
it demonstrates that a remarkably simple fractional deformation of
Newtonian dynamics, governed by a single deformation parameter $\alpha$
and without introducing additional propagating dynamical degrees of
freedom, can generate the essential background ingredients of an
inflationary evolution. Within the same framework, the
model admits a regular pre-inflationary stage that evolves toward a
quasi-de Sitter inflationary attractor, followed by a dynamically
controlled graceful exit and a radiation-like post-inflationary
background. The duration of the accelerated stage depends on the onset
normalization and the endpoint data; once these are specified, it
enters background-level consistency relations involving the
fractional parameter and the effective inflationary equation of state.

More importantly, the present analysis should be viewed as the early universe
sector of the broader fractional cosmological framework established in
Ref.~\cite{rasouli2026emergent}. Together with the previously established matter-dominated,
and late-time accelerating epochs, the present work completes the
background cosmological evolution within a unified fractional framework.
From this perspective, the principal novelty is not only the construction of
a successful inflationary scenario without an inflaton field, but also the
demonstration that a substantial part of the observed cosmological history
can emerge from the same fractional variational principle.

To characterize the early-time inflationary regime, we consider the region where the effective force remains positive, $\mathcal F(a)>0$, and can support accelerated expansion. To obtain a simple analytical description of this phase, we focus on the limit $x \ll 1$, where the force approaches a nearly constant value. Using equation \eqref{K-def}, a straightforward Taylor expansion yields
\begin{eqnarray}\label{K-expan-1}
\mathcal{F} =\mathcal{F}_0\left[1-\left(\frac{p+2}{2}\right)x + \mathcal{O}(x^2)\right].
\end{eqnarray}

In the early times, well before the onset of the radiation-like post-inflationary regime, 
we may use $\mathcal F(x) \simeq \mathcal{F}_0$, which provides the constant force approximation appropriate to the inflationary regime. 
In this regime, the matter density $\rho_m(a)\propto a^{-3}$ is subdominant and the equation \eqref{acc-eq} reduces to
\begin{eqnarray}\label{inf-gen-eq}
\dot{H}+H^2+\gamma_\alpha(t)H-\mathcal{F}_0\approx0. 
\end{eqnarray}
To obtain the leading behavior, we focus on the quasi-static approximation $\lvert\dot{H} \rvert\ll H^2$, which allows us to obtain a dominant inflationary phase. 
Under this approximation, \eqref{inf-gen-eq} yields two branches:
  \begin{eqnarray}\label{inf-gen-H}
H(t)=\frac{1}{2}\left[-\gamma_\alpha(t)\pm \sqrt{\gamma_{\alpha}^2+4\, \mathcal{F}_0}\right].
\end{eqnarray}

Within the algebraic expression \eqref{inf-gen-H}, the condition
$\gamma_\alpha^2 \simeq 4F_0$ defines the scale at which the two
terms under the square root are comparable. We denote this algebraic
balance scale by
\begin{equation}\label{cross-apr-pre}
t_{\rm bal} \simeq \frac{|\alpha-1|}{2\sqrt{\mathcal{F}_0}}.
\end{equation}
This scale characterizes the algebraic approximation \eqref{inf-gen-H} only and
does not define a dynamical transition or the onset of the quasi-de
Sitter stage on the exact regular branch.

For $\gamma_\alpha^2 \ll 4\mathcal{F}_0$, equivalently
$t \gg t_{\rm bal}$, the upper branch of \eqref{inf-gen-H} reduces to
\begin{equation}\label{H-inf}
H_{\rm inf}(t) \approx \sqrt{\mathcal{F}_0}
-\frac{\gamma_\alpha(t)}{2}
= \sqrt{\mathcal{F}_0}-\frac{\alpha-1}{2t},
\end{equation}

corresponds to a regime in which the evolution of the universe is described by a de Sitter-like behavior.

Moreover, the scale factor of this quasi de Sitter regime is given by
\begin{eqnarray}\label{a-inf}
a_{\rm inf}(t)\simeq
a_{\rm on}
\left(\frac{t}{t_i}\right)^{\frac{1-\alpha}{2}}
\exp\!\left[\sqrt{\mathcal{F}_0}\,(t-t_i)\right], 
\end{eqnarray}
where $t_i>0$ denotes a reference time within the quasi-de Sitter
regime, and $a_{\rm on}\equiv a_{\rm inf}(t_i)$ fixes the normalization
of the scale factor. In Subsection~\ref{H-pre}, $t_i$ will be fixed
operationally from the exact regular branch.

The solution~\eqref{a-inf} closely resembles those obtained in
Refs.~\cite{rasouli2014noncommutative, Rasouli:2016syh}, where kinetic inflation was shown to arise, in the
absence of ordinary matter, within a noncommutative Brans--Dicke
framework.

We now derive the remaining fractional quantities associated
with the quasi-de Sitter regime. Substituting the scale factor from
equation~\eqref{a-inf} into equation~\eqref{frac-T}, we obtain the leading-order fractional kinetic
contribution as
\begin{equation}\label{Talpha-late-part}
T_{\alpha}^{(\mathrm{inf})}(t)
\simeq
C_T^{(\mathrm{inf})}(\alpha)
+
\frac{(\alpha-1)\sqrt{\mathcal{F}_0}\,
a_{\mathrm{inf}}^{2}(t)}{2t},
\end{equation}
where $C_T^{(\mathrm{inf})}(\alpha)$ is the additive constant
associated with the inflationary regime. Its value is
fixed by matching to the exact regular branch and by the
adopted energy convention.
In the regime $a\ll a_c$, the fractional potential~\eqref{phi-alpha-def}
reduces to
\begin{equation}
\Phi_{\alpha}^{(\mathrm{inf})}(a)
\simeq
-\frac{\mathcal{F}_0}{2}a^2+C_{\Phi}(\alpha).
\end{equation}
Defining
\begin{equation}
C_{\mathrm{inf}}(\alpha)
\equiv
C_T^{(\mathrm{inf})}(\alpha)+C_{\Phi}(\alpha),
\end{equation}
the fractional energy~\eqref{cos-E-def} becomes
\begin{equation}\label{E-alpha-late-part}
E_{\alpha}^{(\mathrm{inf})}(t)
\simeq
C_{\mathrm{inf}}(\alpha)
-\frac{a_{\mathrm{inf}}^{2}(t)}{2}
\left[
\mathcal{F}_0+
\frac{\sqrt{\mathcal{F}_0}(1-\alpha)}{t}
\right].
\end{equation}

Consequently, the corresponding effective energy density~\eqref{rho-eff} and pressure~\eqref{p-eff} associated with the inflationary regime become
\begin{eqnarray}\label{rho_inf}
\rho_{\alpha}^{(\mathrm{inf})}(t)
&\simeq&
\frac{3}{8\pi G}
\left[
\mathcal{F}_0+
\frac{\sqrt{\mathcal{F}_0}(1-\alpha)}{t}
\right]
-
\frac{3C_{\mathrm{inf}}(\alpha)}
{4\pi G\,a_{\mathrm{inf}}^{2}(t)},
\\\nonumber\\
\label{p_inf}
p_{\alpha}^{(\mathrm{inf})}(t)
&\simeq&
-\frac{3}{8\pi G}
\left[
\mathcal{F}_0+
\frac{\sqrt{\mathcal{F}_0}(1-\alpha)}{t}
\right]
+
\frac{C_{\mathrm{inf}}(\alpha)}
{4\pi G\,a_{\mathrm{inf}}^{2}(t)}.
\end{eqnarray}
The constant $C_{\rm inf}(\alpha)$ is not independent of the
normalization of the exact regular branch. As shown in~\ref{AppendixA}, particularly equation~\eqref{app-Cinf-matching}, matching the regular and
quasi-de Sitter representations gives
$C_{\rm inf}(\alpha)=-\mathcal K/2$. Hence,
$C_{\rm inf}(\alpha)=0$ in the spatially flat model considered
in this work. If retained before imposing $\mathcal K=0$, the corresponding
terms in equations~\eqref{rho_inf} and~\eqref{p_inf} scale as $a_{\rm inf}^{-2}$ and
represent the usual curvature-like contribution. After imposing
spatial flatness, equations~\eqref{rho_inf} and~\eqref{p_inf} reduce to their leading
quasi-de Sitter forms. The first nonvanishing departure of
$w_{\rm eff}^{(\rm inf)}=p_{\rm eff}/\rho_{\rm eff}$ from $-1$
occurs at the next order and is most directly retained through
the Hubble-flow identity given below.

Moreover, employing~\eqref{eff-frd} and~\eqref{acc-eff-eq}, the effective equation of state
(EoS) parameter during the inflationary regime can be written as
\begin{equation}\label{w-H}
w_{\rm eff}^{(\rm inf)}
=
-1-\frac{2}{3}\frac{\dot H_{\rm inf}}{H_{\rm inf}^{2}}
=
-1+\frac{2}{3}\epsilon_{_H}^{(\rm inf)} ,
\end{equation}
where
\begin{equation}\label{epsilon_H_def}
\epsilon_{_H}^{(\rm inf)}
\equiv
-\frac{\dot H_{\rm inf}}{H_{\rm inf}^{2}} .
\end{equation}
This representation is particularly useful because it allows the
inflationary character of the solution to be examined directly from
the background dynamics, without introducing an additional estimate
for the characteristic time scale.

Substituting the inflationary solution \eqref{H-inf} into equation \eqref{epsilon_H_def}
gives
\begin{align}\label{epsilon_H_exact}
\epsilon_{_H}^{(\rm inf)}=
\frac{2(1-\alpha)}
{\left[2\sqrt{\mathcal{F}_0}\,t+(1-\alpha)\right]^2}.
\end{align}
Accordingly, the effective EoS parameter becomes
\begin{equation}\label{weff_exact}
w_{\rm eff}^{(\rm inf)}
=
-1+
\frac{4(1-\alpha)}
{3\left[2\sqrt{ \mathcal{F}_0}\,t+(1-\alpha)\right]^2}.
\end{equation}
Equation~\eqref{weff_exact} is obtained directly from the inflationary
background solution and therefore retains its explicit time dependence.

It is important to interpret the result \eqref{weff_exact} within the same
regime in which the inflationary solution itself was derived. In
obtaining equation \eqref{H-inf}, we assumed that the fractional
correction is subdominant with respect to the approximately constant
effective-force contribution, namely we used $\gamma_\alpha^2(t)\ll 4\mathcal{F}_0 $, or
equivalently,
\begin{equation}\label{alpha_condition}
|\alpha-1|\ll2\sqrt{ \mathcal{F}_0}\,t .
\end{equation}
Notice that the latter condition involves the absolute value
$|\alpha-1|$ and is therefore applicable on both sides of
$\alpha=1$.

Using equation~\eqref{alpha_condition}, equation~\eqref{weff_exact} reduces to
\begin{equation}\label{weff_approx}
w_{\rm eff}^{(\rm inf)}
\simeq
-1+\frac{1-\alpha}{3 \mathcal{F}_0t^2}.
\end{equation}
Interestingly, equation \eqref{weff_approx} also distinguishes the two
sides of the fractional parameter space. For $\alpha<1$, one has $w_{\rm eff}^{(\rm inf)}>-1$
so that the quasi-de Sitter regime is approached from the
$w_{\rm eff}>-1$ side. Conversely, for $\alpha>1$, we get $w_{\rm eff}^{(\rm inf)}<-1$
corresponding to an effective phantom-like approach to the de Sitter
value. These two possibilities should be understood as effective
background behaviors generated by the fractional dynamics rather than
as the introduction of additional quintessence or phantom matter
components.

Furthermore, within the regime \eqref{alpha_condition}, the magnitude of the correction is
suppressed as $t^{-2}$,
\begin{equation}\label{Mag-W}
  \left|1+w_{\rm eff}^{(\rm inf)}\right|
    \simeq
    \frac{|\alpha-1|}
    {3\mathcal{F}_{0}t^{2}}.
\end{equation}
Equation~\eqref{Mag-W} makes explicit that the departure from the de Sitter
value is controlled by the dimensionless combination
$|\alpha-1|/(\mathcal{F}_0t^2)$. Consistently, equation~\eqref{epsilon_H_exact} shows that, at leading
order in the inflationary regime, the quasi-de Sitter condition
$|\epsilon_{_H}|\ll1$ requires
\begin{equation}\label{QdeSitt-con}
|\alpha-1|\ll 2 \mathcal{F}_0t^2.
\end{equation}
Under this condition, $|w_{\rm eff}^{(\rm inf)}+1|\ll1$, and the
inflationary background therefore remains close to the de Sitter
limit, $w_{\rm eff}^{(\rm inf)}\simeq-1$. In the asymptotic regime, where the fractional correction becomes negligible,
the inflationary solution approaches the de Sitter limit,
$H_{\rm inf}\to\sqrt{ \mathcal{F}_0}$, $\epsilon_{_H}\to 0$, and consequently
$w_{\rm eff}^{(\rm inf)}\to -1$.
We stress that these conclusions follow directly from the inflationary
solution and its domain of validity. In particular, no identification
such as $t\sim H_{\rm inf}^{-1}$ is required, and no additional
algebraic bound on $\alpha$ is inferred from the effective EoS parameter.

We should note that two complementary conditions characterize the
regime considered here. (i) The self-consistency condition~~\eqref{alpha_condition}
underlies the approximate inflationary solution and controls the
fractional correction relative to the approximately constant
contribution $\mathcal{F}_0$. (ii) In addition, the quasi-de Sitter
requirement $|\epsilon_{_H}|\ll 1$ yields, at leading order,
condition~\eqref{QdeSitt-con}, ensuring that the Hubble parameter varies sufficiently
slowly and that $w_{\rm eff}^{({\rm inf})}$ remains close to $-1$.
The relative strength of these two conditions depends on the
dimensionless combination $\sqrt{\mathcal{F}_0}\,t$; therefore,
neither condition is in general identified with the other.

The subsequent evolution of the inflationary scenario, including
its duration, and the graceful exit, will be investigated in the next sections using the full background dynamics.

\section{Pre-inflationary regime}
\label{pre-inf}

We now turn to the very early stage preceding the quasi-de Sitter
inflationary regime. In the previous section, the inflationary
solution was obtained under the approximation
$|\dot H|\ll H^2$, which need not hold at sufficiently early times.
We therefore retain the constant-force approximation
$\mathcal{F}(a)\simeq \mathcal F_0$ and the subdominance of ordinary matter, but
solve the reduced background equation~\eqref{inf-gen-eq} exactly, without imposing
the quasi-de Sitter condition.

It is straightforward to show that the general exact solution of
equation~\eqref{inf-gen-eq} can be written as
\begin{equation}\label{a_exact_bessel}
a(t)=
y^\nu
\left[
C_1 I_\nu(y)
+
C_2 K_\nu(y)
\right],
\qquad
y\equiv\sqrt{\mathcal{F}_0}\,t,
\qquad
\nu\equiv1-\frac{\alpha}{2},
\end{equation}
where $I_\nu$ and $K_\nu$ denote the modified Bessel functions of the
first and second kinds, respectively.

It is important to emphasize that equation~\eqref{a_exact_bessel} has been obtained
without imposing $|\dot H|\ll H^2$. It is therefore an exact solution
of the reduced background equation~\eqref{inf-gen-eq}, subject only to the
approximations $\mathcal F(a)\simeq\mathcal F_0$ and the subdominance
of the ordinary matter component adopted above.

\subsection{Early-time behavior and selection of the regular branch}
\label{regular_branch}

For the analysis of the solutions associated with the very early universe in the limit $t\rightarrow0^+$, it is convenient to
rewrite the general solution \eqref{a_exact_bessel} in terms of two linearly independent solutions whose
small-$t$ behavior is explicit. Concretely, an equivalent representation of the
general solution \eqref{a_exact_bessel} is
\begin{eqnarray}\nonumber
a(t)&=\!&a_{_{A}}(t)+a_{_{B}}(t)\nonumber\\\nonumber\\
&=&A\,
{}_0F_1
\left(
;\frac{\alpha}{2};
\frac{\mathcal{F}_0\,t^2}{4}
\right)
+
B\,t^{2-\alpha}
{}_0F_1
\left(
;2-\frac{\alpha}{2};
\frac{\mathcal{F}_0\,t^2}{4}
\right),
\label{a_exact_hyper}
\end{eqnarray}
where ${}_0F_1(;b;z)$ denotes the confluent hypergeometric limit
function,
\begin{equation}\label{0F1}
{}_0F_1(;b;z)
=
\sum_{k=0}^{\infty}
\frac{z^k}{(b)_k\,k!},
\end{equation}
with $(b)_k$ denoting the Pochhammer symbol.

For generic values of $\alpha$ for which the two solutions are
independent, using the expansion
\begin{equation}\label{0F1-expansion}
{}_0F_1(;b;z)
=
1+\frac{z}{b}
+\frac{z^2}{2b(b+1)}
+\mathcal{O}(z^3),
\end{equation}
the first mode, $a_{_{A}}(t)$, behaves as 
\begin{equation}\label{first_mode_expansion}
{}_0F_1
\left(
;\frac{\alpha}{2};
\frac{\mathcal{F}_0\,t^2}{4}
\right)
=
1+
\frac{\mathcal{F}_0\,t^2}{2\alpha}
+
\frac{\mathcal{F}_0^2t^4}
{8\alpha(\alpha+2)}
+
\mathcal{O}(t^6),
\end{equation}
whereas the second mode, $a_{_{B}}(t)$, behaves as
\begin{equation}\label{second_mode_expansion}
t^{2-\alpha}
{}_0F_1
\left(
;2-\frac{\alpha}{2};
\frac{\mathcal{F}_0\,t^2}{4}
\right)
=
t^{2-\alpha}
\left[
1+
\frac{\mathcal{F}_0\,t^2}{2(4-\alpha)}
+
\mathcal{O}(t^4)
\right].
\end{equation}

Consequently, the general early-time behavior of the scale factor in equation \eqref{a_exact_hyper} takes the form
\begin{equation}\label{a_early_general}
a(t)\simeq
A
\left[
1+\frac{\mathcal{F}_0\,t^2}{2\alpha}
+\mathcal{O}(t^4)
\right]
+
B\,t^{2-\alpha}
\left[
1+\frac{\mathcal{F}_0\,t^2}{2(4-\alpha)}
+\mathcal{O}(t^4)
\right].
\end{equation}

Focusing on the interval $0<\alpha<2$, we now investigate the
regularity properties of the two modes in the limit $t\to0^{+}$.

\noindent
(i) First, one can easily show that both modes $a_{_{A}}(t)$ and $a_{_{B}}(t)$ remain finite as
$t\to0^{+}$.

\noindent
(ii) Considering the interval $1<\alpha<2$, it is seen that both $
\dot a_B(t)\sim B(2-\alpha)t^{1-\alpha}$ and 
$\ddot a_B(t)\sim
B(2-\alpha)(1-\alpha)t^{-\alpha}$ diverge as $t\to0^{+}$ whenever $B\neq0$.
By contrast, $\dot a_A(t)\to0$ and
$\ddot a_A(t)\to A\mathcal{F}_0/\alpha$ remain finite.
Thus, requiring regularity of the background quantities selects
$B=0$.

\noindent
(iii) For $0<\alpha<1$, the second mode satisfies
$\dot a_B(t)\to0$, whereas $\ddot a_B(t)$ diverges as
$t\to0^+$. In contrast, both $\dot a_A(t)$ and
$\ddot a_A(t)$ remain finite.

Consequently, throughout the interval $0<\alpha<2$, $\alpha\neq1$,
requiring the background quantities $a$, $\dot a$, and $\ddot a$
to remain finite as $t\to0^{+}$ selects $B=0$. Hence, imposing
regularity of the early-time background selects the first mode $a_{_A}(t)$ in
equation~\eqref{a_exact_hyper}. We shall refer to this solution as the
\emph{regular branch}.

Therefore, the regular branch selected by the early-time regularity
condition can be written as
\begin{equation}\label{a_pre_exact}
a_{\rm reg}(t)=
a_{\rm min}\,
{}_0F_1
\left(
;\frac{\alpha}{2};
\frac{\mathcal{F}_0\,t^2}{4}
\right),
\qquad
a_{\rm min}\equiv a_{\rm reg}(0)>0.
\end{equation}

The ordinary and fractional kinetic contributions, the fractional
potential and energy, and the conserved total energy evaluated along
the exact regular branch are derived in~\ref{AppendixA},
together with their early-time expansions and their relation to the assumption $\mathcal K=0$.

As a special fixed-$\mathcal{F}_0$ limit of the regular branch, taking
$\alpha\to1$ in equation~\eqref{a_pre_exact} gives
\begin{equation}
a_{\rm reg}(t)\longrightarrow
a_{\rm min}\,{}_0F_1\!\left(; \frac{1}{2};\frac{\mathcal{F}_0t^2}{4}\right)
=a_{\rm min}\cosh(\sqrt{\mathcal{F}_0}\,t),
\qquad \mathcal{F}_0>0.
\end{equation}
This special limit illustrates the $\alpha\to1$ behavior of the
reduced solution and does not replace the general pre-inflationary
analysis, which concerns the limit $t\to0^+$ at fixed
$0<\alpha<2$, $\alpha\neq1$.

The very early-time expansion of the regular trajectory \eqref{a_pre_exact}, denoted by $a_{\rm pre}(t)$, reads
\begin{equation}\label{a_pre_expansion}
a_{\rm pre}(t)
=
a_{\rm min}
\left[
1+
\frac{\mathcal{F}_0\,t^2}{2\alpha}
+
\frac{\mathcal{F}_0^2t^4}
{8\alpha(\alpha+2)}
+
\mathcal{O}(t^6)
\right].
\end{equation}

Here, it is important to highlight another key feature of the regular solution \eqref{a_pre_exact}. The equivalence between the hypergeometric and modified-Bessel
representations follows from the identity
\begin{equation}\label{F1-I}
{}_0F_1\left(;b;\frac{y^2}{4}\right)
=
\Gamma(b)
\left(\frac{y}{2}\right)^{1-b}
I_{b-1}(y).
\end{equation}
For the regular solution \eqref{a_pre_exact}, we set  $b=\alpha/2$, while
$\nu=1-\alpha/2$. Hence, $b-1=-\nu$
and therefore \eqref{F1-I} can be rewritten as
\begin{equation}
{}_0F_1\left(;\frac{\alpha}{2};\frac{y^2}{4}\right)
=
\Gamma\left(\frac{\alpha}{2}\right)
\left(\frac{y}{2}\right)^\nu I_{-\nu}(y).
\label{hyper_bessel_relation}
\end{equation}
Consequently, the regular branch selected in
 solution \eqref{a_pre_exact} is equivalently given by
\begin{equation}\label{regular_bessel_form}
a_{\rm reg}(t)
=
a_{\rm min}\Gamma\left(\frac{\alpha}{2}\right)
\left(\frac{y}{2}\right)^\nu I_{-\nu}(y),
\end{equation}
where $y$ and $\nu$ were defined in equation \eqref{a_exact_bessel}.

For non-integer $\nu$, the identity
\begin{equation}\label{Iminus_relation}
I_{-\nu}(y)
=
I_{\nu}(y)
+
\frac{2}{\pi}\sin(\pi\nu)K_{\nu}(y)
\end{equation}
shows explicitly that the regular hypergeometric branch corresponds
to a definite linear combination of the two modified-Bessel modes
appearing in the general solution~\eqref{a_exact_bessel}. Therefore, the
hypergeometric and modified-Bessel representations span the same
two-dimensional solution space. The condition $B=0$ selects a
specific regular trajectory within this space and does not remove,
or subsequently reintroduce, an independent physical mode.

\subsection{Exact Hubble parameter and consistency of the
pre-inflationary dynamics}
\label{H-pre}

The Hubble parameter corresponding to
equation~\eqref{a_pre_exact} can be written as
\begin{equation}\label{H_pre_exact}
H_{\rm reg}(t)\equiv \frac{\dot a_{\rm reg}}{a_{\rm reg}}
=
\frac{\mathcal{F}_0\,t}{\alpha}
\,
\frac{
{}_0F_1
\left(
;1+\frac{\alpha}{2};
\frac{\mathcal{F}_0\,t^2}{4}
\right)
}{
{}_0F_1
\left(
;\frac{\alpha}{2};
\frac{\mathcal{F}_0\,t^2}{4}
\right)
}.
\end{equation}

To define the onset of the quasi-de Sitter stage operationally, we
use the dimensionless time $y$ given by equation \eqref{a_exact_bessel}
and write
\begin{equation}
H_{\rm reg}(t)=\sqrt{\mathcal{F}_0}\,h (y), \qquad 
h(y) \equiv
\frac{y}{\alpha}
\frac{{}_0F_1\!\left(;1+\frac{\alpha}{2};\frac{y^2}{4}\right)}
     {{}_0F_1\!\left(;\frac{\alpha}{2};\frac{y^2}{4}\right)}.
\end{equation}
The Hubble slow-roll parameter is therefore
\begin{equation}\label{H-h}
\epsilon_{_H}^{(\rm reg)}(y;\alpha)
\equiv-\frac{\dot H_{\rm reg}}{H_{\rm reg}^2}
=-\frac{1}{h^2(y)}\frac{d h}{dy}.
\end{equation}
We define $t_i$ as the first time on the regular expanding branch at
which
$\left|\epsilon_H^{(\rm reg)}(y_i;\alpha)\right|=0.1$.
We fix the onset normalization introduced in equation~\eqref{a-inf} by matching
it to the exact regular branch, $a_{\rm on}\equiv a_{\rm reg}(t_i)$. For the representative
values of $\alpha$ considered here, this condition gives
$y_i\simeq1.73,1.80,1.94,2.01,2.28$ for
$\alpha=0.90,0.95,1.05,1.10,1.30$, respectively. This operational
criterion identifies the onset of the quasi-de Sitter stage along the
exact regular branch; it is not a transition time inferred from the
algebraic approximation. 
The corresponding dimensional onset time is
$t_i=y_i/\sqrt{\mathcal F_0}$.

In the very early regime, the regular 
Hubble trajectory~\eqref{H_pre_exact} reduces to the pre-inflationary form as
\begin{equation}\label{H_pre_expansion}
H_{\rm pre}(t)
=
\frac{\mathcal{F}_0}{\alpha}t
-
\frac{\mathcal{F}_0^2}{\alpha^2(\alpha+2)}t^3
+
\mathcal{O}(t^5).
\end{equation}

The exact early-time behavior also makes the domain of validity of the
quasi-de Sitter approximation transparent. From
equation \eqref{H_pre_expansion},
\begin{equation}\label{H_Hdot_pre}
H_{\rm pre}
\simeq
\frac{\mathcal{F}_0}{\alpha}t,
\qquad
\dot H_{\rm pre}
\simeq
\frac{\mathcal{F}_0}{\alpha},
\end{equation}
and therefore
\begin{equation}\label{pre_consistency_ratio}
\frac{|\dot H_{\rm pre}|}{H_{\rm pre}^2}
\simeq
\frac{\alpha}{\mathcal{F}_0\,t^2}.
\end{equation}

For $\mathcal{F}_0 t^2 \ll \alpha$, equation~\eqref{pre_consistency_ratio}
gives ${|\dot{H}_{\rm pre}|}/{H_{\rm pre}^2} \gg 1$, showing explicitly that the deep early-time regime lies outside the
quasi-de Sitter domain. Thus, the quasi-de Sitter approximation is not
self-consistently satisfied in this limit, confirming that the reduced
early-time equation must be solved without imposing
$|\dot{H}| \ll H^2$. As the evolution proceeds, this early-time regime
is left behind and the quasi-de Sitter condition is progressively
recovered as the solution approaches the inflationary branch.

Since $\dot{H}_{\rm pre}\simeq \mathcal F_0/\alpha>0$, the regular expanding
branch is super-accelerating in the sufficiently early time regime.
More explicitly,
\begin{equation}\label{w_pre}
\epsilon_{_H}^{(\rm pre)}\equiv -\frac{|\dot H_{\rm pre}|}{H_{\rm pre}^2}\simeq -\frac{\alpha}{\mathcal F_0 t^2}<0,
\qquad  w_{\rm eff}^{(\rm pre)}
=-1+\frac{2}{3}\epsilon_{_H}^{(\rm pre)}<-1,
\end{equation}
for sufficiently small positive $t$, even though $H\to0$ as
$t\to0^+$.

\subsection{Absence of an early-time curvature singularity}
\label{nonsingularity}

We now examine the geometrical regularity of the 
regular branch. 
As shown in Subsection \ref{regular_branch}, imposing regularity selects the branch \eqref{a_pre_exact}, 
whose early-time expansion is given by equation \eqref{a_pre_expansion}.

Throughout this subsection, whenever a quantity $X(0)$ is written,
it denotes the limiting value of the regular branch, $X(0)\equiv \lim_{t\to0^{+}}X(t)$, provided that this limit exists and is finite. 
It is worth noting that, in the following, the label ``pre'' denotes the early-time
asymptotic form of the exact regular branch. In particular,
\begin{equation}\nonumber
 a_{\rm reg}(t)\sim a_{\rm pre}(t),
\qquad
H_{\rm reg}(t)\sim H_{\rm pre}(t),
\qquad t\to0^+.
\end{equation}
Hence, whenever the corresponding limits exist, the limiting values
obtained from the early-time expansions coincide with those of the
exact regular trajectory.

Before discussing the curvature invariants, it is useful to address
an apparently singular term in the reduced background equation \eqref{inf-gen-eq}.
Although $\gamma_\alpha(t)$ contains an explicit $1/t$
dependence, it does not enter equation~\eqref{inf-gen-eq}
independently, but through the product $\gamma_\alpha H$. For the
regular branch, the early-time behavior is given by equation \eqref{H_pre_expansion}.

Consequently, it is straightforward to show that
\begin{equation}
\lim_{t\to0^{+}}
\gamma_\alpha(t)H_{\rm pre}(t)
=
\frac{\alpha-1}{\alpha}\mathcal{F}_0,
\label{gammaH_limit}
\end{equation}
which is finite. Thus, although $\gamma_\alpha(t)$ is singular at
$t=0$ when considered separately, the combination
$\gamma_\alpha(t)H_{\rm pre}(t)$ entering the background equation
has a finite limit as $t\to0^+$. The explicit $1/t$ dependence therefore does not generate
a divergence at the level of the background equations.

We now examine the geometrical quantities characterizing the
space-time as ${t\to0^{+}}$. 

For $\mathcal{F}_0>0$ and $0<\alpha<2$, with $\alpha\neq1$, using equation~\eqref{H_pre_expansion}, we obtain
\begin{equation}\label{H_initial}
H_{\rm pre}(0)=0,
\qquad
\dot H_{\rm pre}(0)
=
\frac{\mathcal{F}_0}{\alpha}>0.
\end{equation}
Moreover, equation~\eqref{a_pre_expansion} immediately gives
\begin{equation}\label{a_derivatives_origin}
a_{\rm pre}(0)=a_{\rm min}>0,
\qquad
\dot a_{\rm pre}(0)=0,
\qquad
\ddot a_{\rm pre}(0)
=
\frac{a_{\rm min}\mathcal{F}_0}{\alpha}.
\end{equation}
Thus, the regular branch approaches a finite and non-vanishing scale
factor, with a vanishing Hubble expansion rate and a finite acceleration. In particular, as $t\to0^+$,
$a_{\rm pre}\to a_{\rm min}>0$, $\dot a_{\rm pre}\to0$, and
$\ddot a_{\rm pre}\to a_{\rm min}\mathcal F_0/\alpha>0$.

As shown in Section \ref{FNC}, the fractional cosmological equations derived from
the generalized Newtonian action admit an effective relativistic
representation in terms of the Friedmann equations associated with an
FLRW geometry, including the three possible spatial-curvature sectors.
However, in this representation, the matter content is not introduced
by hand as an independently prescribed source on the right-hand side
of the Einstein equations. Rather, the effective cosmological sector
contains both the ordinary matter contribution inherited from standard
Newtonian cosmology and the additional effective contribution emerging
from the fractional dynamics itself. In this sense, the resulting
framework has been considered as an emergent generalized cosmology whose
\textit{effective relativistic description} takes a Friedmann-like form \cite{rasouli2026emergent}.
In the present work, where our purpose is to investigate the dynamics
of the early universe, we restricted the general fractional
cosmological equations to the spatially flat sector. Consequently, the
background evolution studied here admits an effective geometrical
description in terms of a spatially flat FLRW spacetime, with the scale
factor $a(t)$ determined by the fractional dynamics. This effective
geometrical representation allows us to examine directly whether the
regular branch develops a curvature singularity as
$t\to0^+$.

Consequently, we can consider the spatially flat FLRW line element
as the background geometry. Therefore, for the corresponding Ricci scalar
\begin{equation}\label{Ricci_scalar_FRW}
R
=
6\left(\dot H+2H^2\right),
\end{equation}
using equation~\eqref{H_initial}, we therefore obtain
\begin{equation}\label{Ricci_origin}
R(0)
=\frac{6\mathcal{F}_0}{\alpha},
\end{equation}
which is finite.

Moreover, a stronger test is provided by the quadratic Ricci invariant. For a
spatially flat FLRW geometry, we get
\begin{equation}\label{Ricci_squared_FRW}
R_{\mu\nu}R^{\mu\nu}
=
12\left[
\dot H^2
+
3H^2\dot H
+
3H^4
\right].
\end{equation}
Substituting \eqref{H_initial} into \eqref{Ricci_squared_FRW}, we obtain
\begin{equation}
\left.
R_{\mu\nu}R^{\mu\nu}
\right|_{t=0}
=
\frac{12\mathcal{F}_0^2}{\alpha^2},
\label{Ricci_squared_origin}
\end{equation}
which is again finite.

The absence of a curvature divergence can also be verified directly
from the Kretschmann scalar, $\mathcal K_{_{\rm Kr}}\equiv R_{\mu\nu\rho\sigma}R^{\mu\nu\rho\sigma}$. For the spatially flat FLRW metric,
\begin{equation}\label{K_FRW}
\mathcal K_{_{\rm Kr}}
=
12
\left[
\left(\dot H+H^2\right)^2
+
H^4
\right].
\end{equation}
Using \eqref{H_initial}, we obtain
\begin{equation}\label{K_origin}
\mathcal K_{_{\rm Kr}}(0)
=
\lim_{t\to0^{+}}\mathcal K_{_{\rm Kr}}(t)
=
\frac{12\mathcal{F}_0^2}{\alpha^2}
<\infty.
\end{equation}

Finally, the FLRW geometry is conformally flat, so that the Weyl tensor
vanishes identically,
\begin{equation}\label{Weyl_FRW}
C_{\mu\nu\rho\sigma}=0,
\qquad
C_{\mu\nu\rho\sigma}C^{\mu\nu\rho\sigma}=0.
\end{equation}
Collecting the above results, the regular branch
satisfies \eqref{H_initial} and \eqref{a_derivatives_origin}, while its principal curvature invariants obey \eqref{Ricci_origin}, \eqref{Ricci_squared_origin} and \eqref{K_origin}.
All these quantities remain finite as
$t\to0^+$. We therefore conclude that, within the effective FLRW representation, the regular pre-inflationary background is free of an early-time curvature singularity.

\subsection{Asymptotic connection to the inflationary regime}
\label{Assyp-connection}

Within the reduced constant-force dynamics, the large-$y$ behavior of
the exact regular solution~\eqref{a_pre_exact} approaches the inflationary background
obtained in the previous section, thereby connecting the two regimes
along the same trajectory. Here, ``asymptotic'' refers to the
large-$y$ behavior of the reduced solution and not to the late-time
limit of the full cosmological evolution.

According to equation~\eqref{regular_bessel_form}, the
regular branch can be written, up to an overall constant, as $a_{\rm reg}(t)\propto y^\nu I_{-\nu}(y)$.

For $y\gg1$, the modified Bessel function has the asymptotic behavior as
\begin{equation}
I_{-\nu}(y)
\sim
\frac{e^y}{\sqrt{2\pi y}}
\left[
1-\frac{4\nu^2-1}{8y}
+\mathcal O(z^{-2})
\right].
\end{equation}
Consequently, in the asymptotic regime
$\sqrt{\mathcal{F}_0}\,t\gg1$, equation~\eqref{regular_bessel_form} yields
the large-$y$ form of the exact regular trajectory. We denote this
inflationary asymptotic form by $a_{\rm inf}(t)$:
\begin{equation}\label{a-assymp}
a_{\rm inf}(t)
\propto
t^{(1-\alpha)/2}
\exp\!\left(\sqrt{\mathcal F_0}\,t\right)
\left[
1+\mathcal O\!\left(
\frac{1}{\sqrt{\mathcal F_0}\,t}
\right)
\right].
\end{equation}
Correspondingly, the large-time asymptotic form of the Hubble
parameter of the same exact regular trajectory is denoted by
$H_{\rm inf}(t)$ and satisfies
\begin{equation}\label{H-assymp}
H_{\rm inf}(t)=
\sqrt{\mathcal F_0}
-\frac{\alpha-1}{2t}
+\mathcal O(t^{-2}).
\end{equation}

At leading order, the asymptotic expressions 
\eqref{a-assymp} and \eqref{H-assymp} reproduce
the inflationary background obtained in the previous section, namely solutions \eqref{H-inf} and \eqref{a-inf}.
In particular, for $\sqrt{F_0}\,t\gg1$, the difference
$H_{\rm reg}(t)-H_{\rm inf}(t)$ becomes asymptotically negligible
within the reduced dynamics. Here, $H_{\rm inf}$ denotes the inflationary solution 
given by \eqref{H-inf}, while $H_{\rm reg}$ is the exact Hubble parameter of the regular
 branch given by equation \eqref{H_pre_exact}.

Therefore, the solution selected by the early-time regularity
condition does not describe a disconnected pre-inflationary branch.
Rather, the same exact trajectory evolves asymptotically from the
regular pre-inflationary regime toward the quasi-de Sitter inflationary
regime. The dynamical character of this connection, and in particular
the stability of the inflationary background under homogeneous
deviations from this asymptotic trajectory, will be investigated
separately in the following section.

The notation used throughout this analysis may therefore be summarized
schematically as
\begin{equation}\nonumber
\bigl(a_{\rm pre},H_{\rm pre}\bigr)
\;\xleftarrow{\;t\to0^+\;}\;
\bigl(a_{\rm reg},H_{\rm reg}\bigr)
\;\xrightarrow{\;\sqrt{\mathcal F_0}\,t\gg1\;}\;
\bigl(a_{\rm inf},H_{\rm inf}\bigr).
\end{equation}
Here, $(a_{\rm reg},H_{\rm reg})$ denotes the exact regular trajectory,
whereas the labels ``pre'' and ``inf'' denote its very early time limit
and inflationary asymptotic forms, respectively.

\section{Dynamics of the pre-inflationary--inflationary connection}
\label{Dyn-pre-infl}

The asymptotic analysis of the previous section establishes the
connection between the regular pre-inflationary solution and the
quasi-de Sitter regime. We now investigate the dynamical character of
this connection by examining whether neighboring homogeneous
background trajectories approach the same asymptotic inflationary
behavior within the reduced constant-force dynamics.

\subsection{Exact Hubble dynamics and neighboring trajectories}
\label{exact_Hub_dyn}


Within the reduced constant-force regime, the scale factor obeys
\begin{equation}\label{transition_a_equation}
\ddot a
+
\frac{\alpha-1}{t}\dot a
-
\mathcal F_0 a
=
0,
\end{equation}
which is obtained from equation \eqref{inf-gen-eq}.
No quasi-de Sitter approximation is imposed in deriving
equation~\eqref{transition_a_equation}. As shown in Subsection~\ref{H-pre}, this approximation fails in
the deep pre-inflationary regime and is recovered only asymptotically
along the exact regular trajectory within the reduced dynamics

We now examine the dynamical behavior of trajectories neighboring the
regular solution. Since the system is explicitly time dependent, it is
natural to study deviations from the exact regular trajectory
$H_{\rm reg}(t)$ rather than from a time-independent equilibrium point.
Therefore, we write
\begin{equation}\label{H_perturbation}
H(t)
=
H_{\rm reg}(t)+\delta H(t),
\end{equation}
where $\delta H$ measures the difference between a neighboring
background trajectory and the exact regular trajectory.

Substituting equation~\eqref{H_perturbation} into the Riccati equation~\eqref{inf-gen-eq} and using
the fact that $H_{\rm reg}(t)$ is itself an exact solution of the
reduced equation, we obtain
\begin{equation}\label{deltaH_exact}
\delta\dot H
+
\left[
2H_{\rm reg}(t)
+
\frac{\alpha-1}{t}
\right]\delta H
+
(\delta H)^2
=
0,
\end{equation}
which is an exact non-linear equation for the Hubble deviation within the
reduced background dynamics; no linearization has been introduced.

\subsection{Exact Hubble deviation and reduction to the second mode}
\label{exact_delta_solution}

For $\delta H\neq0$, we introduce $u(t)\equiv{1}/{\delta H(t)}.$
Therefore, equation~\eqref{deltaH_exact} becomes
\begin{equation}\label{u_equation}
\dot u+
\frac{\dot \mu}{\mu} u
=
1 , 
\end{equation}
which is a first-order linear equation with an integrating factor $\mu(t)$ that obeys
\begin{equation}\label{mu_equation}
\frac{\dot \mu}{\mu}\equiv-\left[
2H_{\rm reg}(t)
+
\frac{\alpha-1}{t}
\right].
\end{equation}

The corresponding integrating factor is given by
\begin{equation}\label{mu_definition}
\mu(t)
=
\frac{t^{1-\alpha}}
{a_{\rm reg}^{\,2}(t)},
\end{equation}
where an irrelevant multiplicative constant has been set to unity.

Equation \eqref{u_equation} then yields
\begin{equation}\label{mu-u}
\mu(t)u(t)
-
\mu_{\rm ref} u_{\rm ref}
=
\int_{t_{\rm ref}}^{t}\mu(\lambda)\,d\lambda,
\end{equation}
where $t_{\rm ref}>0$ is an arbitrary finite reference time and quantities with
subscript ``ref'' denote their values at $t=t_{\rm ref}$.
Substituting $u=1/\delta H$ into \eqref{mu-u}, the exact non-linear deviation is given by
\begin{equation}\label{deltaH_exact_solution}
\delta H(t)
=
\frac{\mu(t)}
{
\dfrac{\mu_{\rm ref}}{\delta H_{\rm ref}}
+
\displaystyle\int_{t_{\rm ref}}^{t}\mu(\lambda)\,d\lambda
},
\end{equation}
which applies wherever the denominator remains non-zero.

Substituting the exact regular scale factor
\eqref{a_pre_exact} into \eqref{mu_definition}, the integrating factor $\mu(t)$ becomes
\begin{equation}
\mu(t)
=
\frac{
t^{1-\alpha}
}{
a_{\rm min}^2
\left[
{}_0F_1\left(
;\dfrac{\alpha}{2};
\dfrac{\mathcal F_0t^2}{4}
\right)
\right]^2
}.
\label{Q_hypergeometric}
\end{equation}

Equations~\eqref{deltaH_exact_solution} and
\eqref{Q_hypergeometric} therefore give the exact evolution of a finite
Hubble deviation from the regular trajectory within the reduced
background dynamics, on any interval over which the denominator in
equation\eqref{deltaH_exact_solution}  remains nonzero.

The exact non-linear solution~\eqref{deltaH_exact_solution} also admits a direct interpretation in terms of the two linearly independent solutions of the corresponding
equation \eqref{transition_a_equation}.


To construct the second linearly independent mode explicitly, we seek a
second solution of the equation \eqref{transition_a_equation} in the form
\begin{equation}\label{a2-h}
a_2(t)=h(t)\,a_{\rm reg}(t),
\end{equation}
where $a_{\rm reg}(t)$ is given by equation~\eqref{a_pre_exact} and $h(t)$ is a non-constant function of $t$.

Substituting $a_2(t)$ from \eqref{a2-h} into the differential equation \eqref{transition_a_equation}
and using the fact that $a_{\rm reg}(t)$ itself satisfies it, we obtain
\begin{equation}
\dot h(t)
=C_h \mu(t)=
C_h\,
\frac{t^{1-\alpha}}{a_{\rm reg}^{\,2}(t)},
\end{equation}
where we used relation \eqref{mu_definition} and $C_h$ is an integration constant. 
A further integration yields
\begin{equation}\label{h}
h(t)
=
C_0+
C_h
\int^t
\frac{\lambda^{1-\alpha}}
{a_{\rm reg}^{\,2}(\lambda)}
\,d\lambda,
\end{equation}
where $C_0$ is another integration constant. 
Upon substituting equation~\eqref{h} into equation \eqref{a2-h}, the term
proportional to $C_0$ merely reproduces the original solution
$a_{\rm reg}(t)$ and therefore does not generate a new independent
mode. Absorbing the remaining normalization
into the coefficient of the second solution, we may choose
\begin{equation}\label{second_solution_RO}
a_2(t)
=
a_{\rm reg}(t)
\int^t
\frac{\lambda^{1-\alpha}}
{a_{\rm reg}^{\,2}(\lambda)}
\,d\lambda
=
a_{\rm reg}(t)\int^t \mu(\lambda)d\lambda.
\end{equation}
Thus, equation \eqref{second_solution_RO} provides the second linearly independent solution 
generated from the regular branch by reduction of order. To make 
its connection with the second mode of the general solution~\eqref{a_exact_hyper}, we note that, substituting the regular solution \eqref{a_pre_exact} into equation \eqref{second_solution_RO}, we obtain
\begin{equation}
a_2(t)
=
\frac{1}{a_{\rm min}}
{}_0F_1\!\left(
;\frac{\alpha}{2};
\frac{\mathcal{F}_0 t^2}{4}
\right)
\int^t
\frac{\lambda^{1-\alpha}}
{\left[
{}_0F_1\!\left(
;\frac{\alpha}{2};
\frac{\mathcal{F}_0\lambda^2}{4}
\right)
\right]^2}
\,d\lambda .
\end{equation}

Choosing the integration constant so that no additional multiple 
of the regular solution is included, the integral representation 
reduces to
\begin{equation}
a_2(t)
=
\frac{1}{a_{\rm min}(2-\alpha)}
t^{2-\alpha}
{}_0F_1\!\left(
;2-\frac{\alpha}{2};
\frac{\mathcal{F}_0t^2}{4}
\right),
\qquad \alpha\neq 2.
\end{equation}

Therefore, apart from an irrelevant overall normalization that can 
be absorbed into its integration coefficient, the solution obtained 
by reduction of order is precisely the second linearly independent 
mode appearing in equation \eqref{a_exact_hyper}. This confirms explicitly that the two 
constructions span the same two-dimensional solution space.

Accordingly, the general exact solution of the reduced equation~\eqref{transition_a_equation} can be written as
\begin{equation}\label{a_pre-a2}
a(t)
=
A\,a_{\rm reg}(t)+B\,a_2(t).
\end{equation}

Using equation \eqref{second_solution_RO}, equation \eqref{a_pre-a2} may equivalently be expressed as
\begin{equation}
a(t)
=
a_{\rm reg}(t)
\left[
A+B\int^t\mu(\lambda)\,d\lambda
\right].
\end{equation}

The corresponding difference between the Hubble parameter of the
general trajectory and that of the regular branch is
\begin{align}
H-H_{\rm reg}
&=
\frac{d}{dt}
\ln\left(\frac{a}{a_{\rm reg}}\right)
\nonumber\\
&=
\frac{d}{dt}
\ln\left[
A+B\int^t \mu(\lambda)d\lambda
\right],
\end{align}
which yields
\begin{equation}\label{deltaH_AB}
H(t)-H_{\rm reg}(t)
=
\frac{
B\,\mu(t)
}{
A+B\displaystyle\int^t\mu(\lambda)d\lambda
}.
\end{equation}

Equation~\eqref{deltaH_AB} has exactly the same structure as the nonlinear Hubble
deviation obtained in equation~\eqref{deltaH_exact_solution}. Thus, $\delta H$ does not
represent an additional dynamical degree of freedom; rather, it
measures the effect of changing the relative weight of the two
linearly independent background modes. It is important to distinguish
this forward-stability analysis from the early-time regularity
selection. Regularity as $t\to0^+$ selects $B=0$, as established in
Subsection~\ref{regular_branch}, whereas the present analysis considers neighboring
trajectories specified at a finite reference time $t_{\rm ref}>0$, for
which a small nonzero admixture of the second mode may consistently be
allowed.

\subsection{Asymptotic attraction and basin of regular forward evolution}
\label{transition_basin}

We now examine the asymptotic behavior of equation~\eqref{deltaH_exact_solution}. As established in Subsection~\ref{Assyp-connection}, the exact regular
trajectory approaches the inflationary asymptotic form given by
equation~\eqref{a-assymp} when $\sqrt{\mathcal F_0}\,t\gg1$. Therefore, to determine
the asymptotic behavior of the integrating factor $\mu(t)$ defined in
equation~\eqref{mu_definition}, we may consistently use this large-time form of the
regular trajectory. Therefore, substituting equation \eqref{a-assymp} into equation \eqref{mu_definition}, we obtain
\begin{align}
\mu(t)
\sim
\frac{1}{C_a^2}
\exp\left(
-2\sqrt{\mathcal F_0}\,t
\right),\qquad
\sqrt{\mathcal F_0}\,t\gg1,
\label{Q_late}
\end{align}
where $C_a$ is a positive constant.

An important consequence is that
\begin{equation}\label{J-0-infinity}
\mathcal{J}_{\infty}(t_{\rm ref})
\equiv
\int_{t_{\rm ref}}^{\infty}\mu(\lambda)\,d\lambda,
\end{equation}
is finite.

Provided that the denominator of equation~\eqref{deltaH_exact_solution} does not vanish at a
finite time, its integral contribution approaches a finite non-zero
constant in the asymptotic regime, whereas the numerator $\mu(t)$
decays exponentially. Consequently, within the asymptotic inflationary
regime, $\sqrt{\mathcal F_0}\,t\gg1$,
\begin{equation}\label{deltaH_late_decay}
\delta H(t)
=
H(t)-H_{\rm reg}(t)
\propto
\exp\left(
-2\sqrt{\mathcal F_0}\,t
\right)
\longrightarrow0 .
\end{equation}

Moreover, according to equation~\eqref{H-assymp}, the exact regular trajectory
itself approaches the inflationary asymptotic solution,
$H_{\rm reg}(t)-H_{\rm inf}(t)\to0$. Combining this result with
equation~\eqref{deltaH_late_decay} shows that neighboring admissible expanding trajectories
approach the same quasi-de Sitter inflationary behavior.

In particular,
\begin{equation}
H(t)-H_{\rm inf}(t)
\longrightarrow0,
\qquad
\sqrt{\mathcal F_0}\,t \gg1 ,
\label{H_to_inflation}
\end{equation}
where $H_{\rm inf}(t)$ denotes the inflationary background derived in
Section~\ref{inflation}. Here, the asymptotic statement refers to the reduced
constant-force inflationary regime and should not be interpreted as the
late-time limit of the full cosmological evolution, for which the
approximation $\mathcal F(a)\simeq\mathcal F_0$ eventually ceases to
apply.

The exact expression \eqref{deltaH_exact_solution} also makes clear that asymptotic attraction does
not require every mathematically possible initial condition to generate a
regular forward trajectory. The relevant
condition is that
\begin{equation}\label{Q_definition}
\mathcal{Q}(t)
\equiv
\frac{\mu(t_{\rm ref})}{\delta H(t_{\rm ref})}
+
\int_{t_{\rm ref}}^{t}\mu(\lambda)d\lambda
\end{equation}
does not vanish during the forward evolution.

Since $\mu(t)>0$ for the regular expanding solution, the integral in
equation~\eqref{Q_definition} is monotonically increasing.

For $\delta H(t_{\rm ref})<0$, the initial term in equation~\eqref{Q_definition}
is negative. Regular forward evolution is ensured when
\begin{equation}\label{basin_condition}
 \left|\delta H(t_{\rm ref})\right|
 <
 \frac{\mu(t_{\rm ref})}{J_\infty(t_{\rm ref})}.
\end{equation}
The saturation of inequality~\eqref{basin_condition} defines the
candidate boundary separating regular forward evolution from the
singular branch. As shown below, this boundary is the exact basin
boundary of the reduced constant-force dynamics.

To establish this result analytically, we retain the previously
defined index $\nu\equiv 1-\frac{\alpha}{2}$ and $y_{\rm ref}\equiv \sqrt{\mathcal F_0}\,t_{\rm ref}$. Then, the regular solution is rewritten, up to an irrelevant normalization constant, as $a_{\rm reg}(t)\propto t^{\nu}I_{-\nu}(\sqrt{\mathcal F_0}t)$, so that $\mu(t)\propto {1}/[{t\,I_{-\nu}^{\,2}(\sqrt{\mathcal F_0}t)}]$.

Using $K_{-\nu}(y)=K_{\nu}(y)$ together with the Wronskian identity
\begin{equation}\label{Wronskian}
\frac{d}{dy}
\left[\frac{K_{-\nu}(y)}{I_{-\nu}(y)}\right]
=
-\frac{1}{y I_{-\nu}^{\,2}(y)},
\end{equation}
the integral in equation~\eqref{J-0-infinity} can be evaluated exactly,
yielding
\begin{equation}\label{mu-J-K}
\frac{\mu(t_{\rm ref})}{J_\infty(t_{\rm ref})}
=
\frac{1}{
t_{\rm ref}
I_{-\nu}(y_{\rm ref})
K_{\nu}(y_{\rm ref})
}.
\end{equation}

Saturation of inequality~\eqref{basin_condition} on the
$\delta H(t_{\rm ref})<0$ side, namely
$\delta H(t_{\rm ref})
=-\mu(t_{\rm ref})/J_\infty(t_{\rm ref})$,
selects the separatrix
$
a_{-}(t)\propto t^\nu K_\nu(\sqrt{\mathcal F_0}t)
$,
whose Hubble parameter is
\begin{equation}\label{H-minus}
H_{-}(t)
=
-\sqrt{\mathcal F_0}\,\frac{K_{1-\nu}(\sqrt{\mathcal F_0}t)}{K_\nu(\sqrt{\mathcal F_0}t)}
<0,
\qquad
H_{-}(t)\longrightarrow -\sqrt{\mathcal F_0}
\quad (t\longrightarrow\infty).
\end{equation}

Initial deviations satisfying
$\delta H(t_{\rm ref})>-\mu(t_{\rm ref})/J_\infty(t_{\rm ref})$
generate regular forward trajectories that
asymptotically approach the expanding attractor.
For deviations below this threshold, the
denominator of equation~\eqref{deltaH_exact_solution} vanishes at a finite time.
Since $H_-(t_{\rm ref})<0$, every initially expanding
trajectory belongs to the  basin of attraction.
Moreover, the large-$y_{\rm ref}$ asymptotics give
\begin{equation}\label{mu-J--t-ref}
\frac{\mu(t_{\rm ref})}{J_\infty(t_{\rm ref})}
\longrightarrow 2\sqrt{\mathcal F_0},
\qquad y_{\rm ref}\gg1.
\end{equation}
These results establish the exact basin boundary
within the reduced constant-force dynamics;
they do not constitute a stability result for
the subsequent evolution under the full
fractional potential.

We stress that the stability statement established here is a
background-level result. It concerns homogeneous deviations between
solutions of the reduced background equation and does not constitute
a stability analysis of scalar or tensor cosmological perturbations.

\subsection{Dynamical interpretation}

Within the reduced constant-force dynamics, the preceding results
show that the exact regular trajectory evolves continuously from the
pre-inflationary regime toward the quasi-de Sitter inflationary
regime, while admissible neighboring homogeneous trajectories
asymptotically approach the same inflationary behavior. This
connection does not require matching two independent background
solutions.

The dynamical sequence may therefore be summarized schematically as
\[
\text{regular pre-inflationary regime}
\quad \longrightarrow \quad
\text{quasi-de Sitter inflationary regime}.
\]

In this background sense, the quasi-de Sitter inflationary behavior
acts as an asymptotic attractor within the admissible expanding sector
of the reduced dynamics.

\section{Post-inflationary regime}
\label{Post-inf}

In this section, we investigate the evolution from the inflationary
phase into the post-inflationary regime by analyzing the effective
force $\mathcal{F}(a)$ near the inflationary exit and in the large-$a$
domain. For the particular choice $p=4$, the fractional background
dynamics admits a radiation-like expansion, $a(t)\propto t^{1/2}$,
over a finite interval.

We further examine the dynamical exit from inflation and derive the
corresponding background-level consistency relations involving the
number of e-folds, the fractional parameter, and the effective
inflationary equation of state. In Subsection~\ref{full-radiation-stability}, the asymptotic
treatment is complemented by the complete-background evolution and
an analysis of neighboring homogeneous trajectories, including the
local contraction at the entrance and the relative growing mode that
controls the duration of the radiation-like interval.

We first identify the characteristic scale $a_*$ at which the fractional 
force vanishes, $\mathcal{F}(a_*)=0$. Using equation~\eqref{K-def},
we obtain
\begin{eqnarray}\label{x-star}
x_*=\left(\frac{a_*}{a_c}\right)^{p}=\left(\frac{2}{p-2}\right), \hspace{10mm}a_*=a_c
\left(\frac{2}{p-2}\right)^{\frac{1}{p}},
\end{eqnarray}
which implies that the exponent $p$ must satisfy $p>2$. 
Accordingly, we restrict our analysis to this physically consistent case.

Using \eqref{K-def}, in the regime $x= (\frac{a}{a_c})^p\gg1$, one can easily show that $\mathcal{F}(a)$ reduces to 
\begin{eqnarray}\label{K-expan-2}
\mathcal{F}(a) \simeq-\mathcal{F}_{r} \left(\frac{a_c}{a}\right)^p, \qquad \mathcal{F}_{r}\equiv\left(\frac{p-2}{2}\right)\mathcal{F}_{0}>0.
\end{eqnarray}

In the post-inflationary regime considered here, we assume that the
ordinary matter contribution remains subdominant relative to the
fractional term. 
Therefore, substituting the asymptotic force~\eqref{K-expan-2} into the reduced
equation~\eqref{acc-eq} gives the large-$a$ power-law solution

\begin{eqnarray}\label{a-rad}
a(t) =a_{r}\, t^{{2}/{p}}, \qquad a_r\equiv a_c\left[-\frac{p^2\,\mathcal{F}_{r} }{2(\alpha-2)p+4}\right]^{{1}/{p}}.
\end{eqnarray}
Requiring the amplitude $a_r$ to be real and positive imposes 
$\alpha<2-{2}/{p}$.
Since $p>2$, this condition provides a stronger upper bound than
$\alpha<2$ and is therefore fully consistent with the range $0<\alpha<2$ adopted in the preceding analysis.

In deriving the above solution, we focus on an early
post-inflationary regime in which the fractional contribution
dominates over the ordinary pressureless matter component. For $p>3$, the fractional contribution
decreases faster than the dust term, so that, whenever the fractional
sector initially dominates, this approximation can hold only over a
finite post-inflationary interval, as quantified below. The particular
choice $p=4$, considered in the next subsection, is of special physical
interest because it yields the standard radiation-like expansion
$a(t)\propto t^{1/2}$.

Using the large-$a$ behavior in equation \eqref{K-expan-2}, together with
$\rho_m(a)=\rho_{m,c}(a_c/a)^3$, where
$\rho_{m,c}\equiv\rho_m(a_c)$, neglecting the ordinary matter term
requires
\begin{equation}\label{mat_dom_gen}
\frac{4\pi G}{3}\rho_m(a)
\ll
\mathcal F_r
\left(\frac{a_c}{a}\right)^p,
\end{equation}
or, equivalently,
\begin{equation}\label{rad_dom_con}
\frac{4\pi G\rho_{m,c}}
{3\mathcal F_r}
\left(\frac{a}{a_c}\right)^{p-3}
\ll 1.
\end{equation}
Thus, the neglect of the matter contribution is not a consequence of
$p>3$ alone, but is controlled by the above dominance condition.
Since, for $p>3$, the ratio of the matter term to the fractional
contribution grows as $a^{p-3}$, the ordinary matter component
eventually becomes dynamically relevant as the universe expands.
This behavior is also consistent with the subsequent emergence of the
matter-dominated regime obtained within the broader fractional
cosmological framework~\cite{rasouli2026emergent,rasouli2026evolution}.

\subsection{\textbf{Special case: $p=4$}}
\label{p_equal_4}
We now examine the above solutions for the particular choice $p=4$. In this case, the condition determining the zero of the fractional force (see equation \eqref{x-star}) yields $x_*=1$, or equivalently $a_*=a_c$. Thus, $\mathcal{F}(a)$ vanishes at $a=a_c$ and changes
sign for $a>a_c$, becoming negative thereafter. This sign reversal
marks the characteristic scale around which the transition from the
inflationary to the post-inflationary dynamics takes place.  As shown
in the following subsection, however, the end of accelerated expansion
is determined kinematically by $\ddot a(t_{\rm end})=0$ and therefore
occurs close to, but not necessarily exactly at $a_*=a_c$.

Furthermore, equation \eqref{K-expan-2} reduces to 
\begin{eqnarray}\label{K-expan-rad}
\mathcal{F}(a) =-\mathcal{F}_{r} \left(\frac{a_c}{a}\right)^4, \qquad \mathcal{F}_{r}\equiv\mathcal{F}_{0}>0.
\end{eqnarray}

Therefore, in the large-$a$ regime and under the matter-subdominance
condition~\eqref{rad_dom_con}, the reduced background equation admits the
radiation-like power-law solution
\begin{eqnarray}\label{a-rad-real}
a_{\rm rad}(t) =a_{r}\, t^{{1}/{2}}, \qquad a_r\equiv a_c\left(\frac{4\,\mathcal{F}_{0} }{3-2\alpha}\right)^{{1}/{4}},
\end{eqnarray}
where we used equation \eqref{a-rad}. The above equation implies a theoretical upper bound on the fractional parameter $\alpha$ as $\alpha < \frac{3}{2}$. Moreover, the correspondence with a radiation-dominated background
is not merely kinematical. Substituting $a(t)$ from equation \eqref{a-rad-real} into the effective
EoS~$w_{\rm eff}^{(\rm rad)}
=
-1-{(2\dot H_{\rm rad})}/{(3H_{\rm rad}^{2})}
=
-1+{(2\epsilon_{_H}^{(\rm rad)})}/{3}$, we obtain
$w_{\rm eff}^{(\rm rad)}=1/3$. Thus, the post-inflationary background reproduces both the expansion
law $a(t)\propto t^{1/2}$ and the effective EoS
$w_{\rm eff}=1/3$ characteristic of a radiation-like FLRW background.

It is worth noting that, while the time dependence
$a(t)\propto t^{1/2}$ coincides with the standard radiation-dominated
behavior in the $\Lambda$CDM scenario, the normalization of the solution
remains explicitly dependent on the fractional parameter $\alpha$.
Thus, the fractional modification leaves a residual imprint on the
normalization of the post-inflationary background solution and could potentially be constrained by cosmological observations.

Direct substitution of \eqref{K-expan-rad} and \eqref{a-rad-real} into the acceleration
equation \eqref{acc-eq}, after consistently neglecting the matter contribution
under the dominance condition \eqref{rad_dom_con}, confirms that \eqref{a-rad-real} is an
exact solution of the resulting post-inflationary equation.

In summary, for the special case $p=4$, the post-inflationary dynamics
admits, over a finite interval, radiation-like background evolution,
$a(t)\propto t^{1/2}$ and $w_{\rm eff}=1/3$. This result concerns the
background expansion only and should not be interpreted as a microscopic
mechanism for reheating or thermalization.

The connection of the inflationary trajectory to the large-$a$
radiation-like branch~\eqref{a-rad-real}, together with the duration of the
resulting radiation-like interval, is examined using the complete
background dynamics in Subsection~\ref{full-radiation-stability}.


\subsection{A controlled exit from inflation and consistency relations}
\label{smooth_exit}

In this subsection, we examine the end of accelerated expansion
near the zero of the fractional force, $a=a_*$. We parameterize
the exit displacement by $\delta a=a_{\rm end}-a_*$ and determine
the conditions under which a local expansion is controlled.
The resulting dynamical balance provides consistency relations
among the exit displacement, the fractional parameter, and the
duration of inflation.

The end of accelerated expansion is defined kinematically by $\ddot { a}_{\rm end}\equiv \ddot a(t_{\rm end})=0$. 
Substituting this condition into equation~\eqref{acc-eq} is equivalent to the dynamical balance
\begin{equation}\label{aend_balance}
a_{\rm end}\,\mathcal F(a_{\rm end})=\gamma_{\alpha}^{\rm end}\,\dot a_{\rm end},
\qquad
\gamma^{\rm end}_{\alpha}\equiv \gamma_{\alpha}(t_{\rm end}),
\qquad
\dot a_{\rm end}\equiv \dot a(t_{\rm end}),
\end{equation}
where the ordinary matter contribution has been neglected consistently
with the dominance condition~\eqref{rad_dom_con}. Unlike $a_*$, the quantity $a_{\rm end}$ depends on the time sector through $\gamma_{\alpha}(t)$ and on the instantaneous expansion rate $\dot a(t)$. Equivalently, the exact end of accelerated expansion satisfies
\begin{equation}\label{aend_w}
\epsilon_H(t_{\rm end})=1,
\qquad
w_{\rm eff}(t_{\rm end})=-\frac{1}{3}.
\end{equation}

Equation \eqref{aend_balance} already shows that $a_{\rm end}$ does not generically coincide with $a_*$: even if $\mathcal{F}(a)$ vanishes at $a_*$, the friction/anti-friction term generally does not vanish there, and therefore $\ddot a$ need not vanish at $a=a_*$. Hence, it is not mathematically consistent to define the end of inflation by $\mathcal{F}(a_*)=0$; the physically correct definition is \eqref{aend_balance}.

We introduce the dimensionless variable $\zeta\equiv a/a_c$, which is
related to the variable $x$ defined in equation~~\eqref{K-def} through
$x=\zeta^p$. For $p=4$, we write
$a_{\rm end}=a_*+\delta a$ and introduce the relative exit displacement
$\varepsilon_{\rm ex}\equiv\delta a/a_*$, so that
$\zeta_{\rm end}\equiv a_{\rm end}/a_c=1+\varepsilon_{\rm ex}$.
Since $a_*=a_c$, the exact effective force~\eqref{K-def} can be written as
\begin{equation} \label{G_y}
      G(\zeta)\equiv \frac{\mathcal{F}(a)}{\mathcal{F}_{0}}
=
    \frac{1-\zeta^{4}}{(1+\zeta^{4})^{2}}.
   \end{equation}
Expanding around the simple zero $\zeta=1$ gives
\begin{equation}
    G(1+\varepsilon_{\rm ex})
    =
    -\varepsilon_{\rm ex}
    +\frac{5}{2}\varepsilon_{\rm ex}^{2}
    -\varepsilon_{\rm ex}^{3}
    +\mathcal{O}(\varepsilon_{\rm ex}^{4}).
    \label{G_taylor}
\end{equation}
Therefore, the leading-order treatment of the exit corresponds to retaining
the linear term in equation~\eqref{G_taylor}. Rather than simply assuming
$|\varepsilon_{\rm ex}|\ll1$, we quantify the accuracy of this truncation.
Let $\eta$ denote the maximum allowed magnitude of the quadratic correction
relative to the retained linear contribution. We therefore impose
\begin{equation}
    \frac{
    \left|\frac{5}{2}\varepsilon_{\rm ex}^{2}\right|
    }{
    |\varepsilon_{\rm ex}|
    }
    =
    \frac{5}{2}|\varepsilon_{\rm ex}|
    \leq \eta,
    \qquad 0<\eta\ll1.
    \label{eta_definition}
\end{equation}
It follows immediately that
\begin{equation}    \label{exit_eta_bound}
    \frac{|\delta a|}{a_{*}}
    =
    |\varepsilon_{\rm ex}|
    \leq
    \frac{2\eta}{5}.
\end{equation}
Equation~\eqref{exit_eta_bound} quantifies the validity of the linearized exit
analysis for a prescribed tolerance $\eta$.

For general $p$, we similarly write $a_{\rm end}=a_*+\delta a$
and expand the force around $a_*$:
\begin{equation}
\mathcal{F}(a_{\rm end})=\mathcal{F}(a_*+\delta a)
=
\mathcal{F}(a_*)+\mathcal{F}_1^*\,\delta a+\frac{1}{2} \mathcal{F}_2^*(\delta a)^2+\mathcal{O}\bigl((\delta a)^3\bigr),
\label{F_expand_astar}
\end{equation}
where
\begin{equation}\label{F1}
\mathcal{F}_1^*\equiv \left.\frac{d\mathcal{F}}{da}\right|_{a_*}=-\frac{\mathcal{F}_0 (p-2)^{2+1/p}}{2^{1/p} a_{\rm c}\, p} ,
\end{equation}
\begin{equation}\label{F2}
\mathcal{F}_2^*\equiv \left.\frac{d^2\mathcal{F}}{da^2}\right|_{a_*}=-\frac{\mathcal{F}_0 (p-9) (p-2)^{\frac{2 (p+1)}{p}}}{4^{1/p} a_{\rm c}^2 \, p}.
\end{equation}
In deriving relations \eqref{F1} and \eqref{F2}, we have used equations \eqref{K-def} and \eqref{x-star}.
Since $\mathcal{F}(a_*)=0$ by definition, retaining the linear term in local expansion gives $\mathcal{F}(a_{\rm end})\simeq \mathcal{F}_1^*\,\delta a$. Substituting this result into \eqref{aend_balance}, we obtain
\begin{equation}\label{deltaa_general}
\delta a \simeq \,\frac{\gamma_{\alpha}^{\rm end}\,H_{\rm end}}{\mathcal{F}_1^*},
\end{equation}
which indicates that the separation between $a_*$ and $a_{\rm end}$ is dynamically controlled by the ratio of the friction term to the force gradient at $a_*$.

Let us rewrite the relation \eqref{deltaa_general} in the following equivalent form 
\begin{equation}\label{deltaa_general-use}
\frac{\delta a}{a_*} \simeq \,\frac{\gamma_{\alpha}^{\rm end} H_{\rm end}}{a_* {\mathcal{F}_1^*}}.
\end{equation}

To relate the exit analysis to the duration of the preceding
quasi-de Sitter stage, we first evaluate the number of e-folds
accumulated along the exact regular branch of the reduced background
derived in Section~\ref{pre-inf}. This provides the inflationary-time
parametrization that will subsequently be matched to the exit
condition~\eqref{deltaa_general-use} .

The number of e-folds accumulated from the operationally defined
onset of the quasi-de Sitter stage at $t_i$ up to a generic time $t$
is evaluated along the exact regular branch as
\begin{equation}\label{N}
N_{\rm on}(t)
\equiv N(t_i,t)
=
\int_{t_i}^{t}H_{\rm reg}(t')\,dt'
=
\ln\left[\frac{a_{\rm reg}(t)}{a_{\rm reg}(t_i)}\right]
=
\ln\left[\frac{a_{\rm reg}(t)}{a_{\rm on}}\right].
\end{equation}

Using the exact regular solution \eqref{a_pre_exact}, the number of e-folds becomes
\begin{equation}\label{N_exact_reg}
N_{\rm on}(t)
=
\ln\!\left[
\frac{{}_0F_1\!\left(;\frac{\alpha}{2};
                      \frac{\mathcal{F}_0t^2}{4}\right)}
     {{}_0F_1\!\left(;\frac{\alpha}{2};
                      \frac{\mathcal{F}_0t_i^2}{4}\right)}
\right].
\end{equation}

In the large-$y$ quasi-de Sitter regime, we expand the exact regular
solution at the evaluation time while retaining the exact
normalization at the onset. 
To retain the onset normalization without requiring $y_i\gg1$,
we define
\begin{equation}
\Delta_{\rm on}\equiv
\ln\left[
\frac{
2^{\alpha/2-1}\Gamma(\alpha/2)\,
e^{y_i}y_i^{(1-\alpha)/2}
}{
\sqrt{2\pi}\,
{}_0F_1\left(;\frac{\alpha}{2};\frac{y_i^2}{4}\right)
}
\right],
\end{equation}
where the initial time $t_i$ is not arbitrary. Throughout this section, it
denotes the operational onset time defined from the exact regular
branch in Subsection~\ref{H-pre} by the condition $|\epsilon_H^{\rm (reg)}(y_i;\alpha)|=0.1$.
The corresponding numerical values of $y_i$ are given below
equation \eqref{H-h}. In particular, $t_i$ is not identified with the algebraic
balance scale introduced in equation \eqref{cross-apr-pre}.

The large-$y$ expansion of equation~\eqref{N_exact_reg} then gives
\begin{equation}\label{N_exact_qdS}
 N_{\rm on}(t)
 \simeq
 \sqrt{\mathcal F_0}\,(t-t_i)
 -\frac{\alpha-1}{2}\ln\left(\frac{t}{t_i}\right)
 +\Delta_{\rm on}.
\end{equation}
For each fixed admissible $\alpha$, $\Delta_{\rm on}$ is a constant
determined by the operational onset. Equivalently, in dimensionless
variables,
\begin{equation}\label{N_z_alpha}
 N_{\rm on}(y)\equiv N(y_i,y)
 \simeq
 y-y_i
 -\frac{\alpha-1}{2}\ln\left(\frac{y}{y_i}\right)
 +\Delta_{\rm on}.
\end{equation}

Within the large-$y$ regime, equation~\eqref{N_z_alpha} relates the accumulated
number of inflationary e-folds to the dimensionless times $y_i$ and
$y$, and to the fractional parameter $\alpha$. This relation does not
by itself fix a unique numerical value of the total e-fold number $N_{\rm tot}$; such a value is obtained
only after specifying the endpoint, for example $y=y_{\rm end}$.

For each fixed admissible value of $\alpha$, the operationally defined
onset value $y_i$ is finite. Dividing equation~\eqref{N_z_alpha} by $y$ gives
\begin{equation}\label{N_y-yi}
\frac{N_{\rm on}(y)}{y}
\simeq
1-\frac{y_i}{y}
-\frac{\alpha-1}{2y}\ln\left(\frac{y}{y_i}\right)
+\frac{\Delta_{\rm on}}{y}.
\end{equation}
Hence, for fixed admissible $\alpha$ and sufficiently large $y$,
all three correction terms vanish, so that
$N_{\rm on}(y)/y\to1$. At leading order, we therefore obtain
\begin{equation}\label{N_approx_z}
N_{\rm on}(t)\simeq y=\sqrt{\mathcal{F}_0}\,t.
\end{equation}

We emphasize that equation~\eqref{N_approx_z} is not an independent assumption
concerning the duration of inflation. It follows as the leading-order
limit of the asymptotic relation~\eqref{N_z_alpha}. The subleading form~\eqref{N_z_alpha} may
be retained whenever the onset and logarithmic corrections are
numerically relevant, whereas the exact relation~\eqref{N_exact_reg} should be used
outside the asymptotic domain.

Within the large-$y$ quasi-de Sitter regime, the asymptotic EoS
relation~\eqref{Mag-W} remains applicable. Combining equation~\eqref{Mag-W} with the leading-order relation  $y\simeq N_{\rm on}(t)$ from equation~~\eqref{N_approx_z} gives
\begin{equation}\label{w_alpha_N_final}
\left|1+w_{\rm eff}^{(\rm inf)}(t)\right|
\simeq \frac{|\alpha-1|}{3N_{\rm on}^2(t)}.
\end{equation}
where $w_{\rm eff}^{(\rm inf)}$ denotes the effective EoS
parameter of the inflationary background, here evaluated within the
quasi-de Sitter regime. Equation~\eqref{w_alpha_N_final} represents only its asymptotic
quasi-de Sitter form and should not be extrapolated to the end of
inflation, where $w_{\rm eff}(t_{\rm end})=-1/3$; the subleading onset and logarithmic corrections are
retained in equation~\eqref{N_z_alpha}.

Equivalently, the relation~\eqref{w_alpha_N_final} can be rewritten as
\begin{equation}\label{N_alpha_w_final}
N_{\rm on}(t)
\simeq
\sqrt{
\frac{|\alpha-1|}
{3\left|1+w_{\rm eff}^{(\rm inf)}(t)\right|}
},
\end{equation}
which establishes a same-epoch quasi-de Sitter consistency
relation among $N_{\rm on}(t)$, $\alpha$, and
$w_{\rm eff}^{(\rm inf)}(t)$. It does not by itself determine the
total duration of inflation.

At the end of inflation, we denote the total number of e-folds
accumulated from the operational onset by
\begin{equation}\label{N-tot}
N_{\rm tot}\equiv \ln\left(\frac{a_{\rm end}}{a_{\rm on}}\right),
\end{equation}
which is fixed only after the onset normalization and the
endpoint are specified. Using equation~\eqref{a_pre_exact}, one has
$a_{\rm on}=a_{\min}\mathcal{A}(y_i;\alpha)\equiv a_{\min}{}_0F_1\left(
;\frac{\alpha}{2};
\frac{\mathcal{F}_{0}t_i^2}{4}
\right)$. Therefore,
$N_{\rm tot}$ depends explicitly on the integration constant
$a_{\min}$ and is not determined by $\alpha$ or by the force
parameters alone. In particular, for $a_{\rm end}\simeq a_c$,
one obtains schematically
$
N_{\rm tot}\simeq
\ln({a_c}/{a_{\min}})
-\ln\mathcal{A}(y_i;\alpha),
$
up to the finite onset and exit corrections.

The time, Hubble-flow, EoS parameter, and e-fold conventions used in this section are summarized in ~\ref{AppendixB}.

We now focus on the physically relevant case considered in this work,
namely the special choice $p=4$. In this case, using relations~~\eqref{x-star}
and~ \eqref{F1}, we obtain $a_{*}\mathcal{F}_{1}^{*}=-\mathcal{F}_{0}$
Substituting this result and
$|\gamma_{\alpha}^{\rm end}|=|\alpha-1|/t_{\rm end}$ into
equation~\eqref{deltaa_general-use}, while keeping the exit value of the Hubble parameter
explicit, gives
\begin{equation}\label{a-N}
\frac{|\delta a|}{a_{*}}
\simeq
\frac{|\alpha-1|H_{\rm end}}
     {\mathcal{F}_{0}t_{\rm end}}
=
\frac{|\alpha-1|h_{\rm end}}{y_{\rm end}},
\qquad
h_{\rm end}\equiv h_{\alpha}(t_{\rm end})\equiv
\frac{H_{\rm end}}{\sqrt{\mathcal{F}_{0}}},
\qquad
y_{\rm end}\equiv
\sqrt{\mathcal{F}_{0}}\,t_{\rm end}.
\end{equation}
No specific value of $h_{\rm end}$ is assumed here; the kinematic
condition $\epsilon_{H}(t_{\rm end})=1$ fixes $\dot H_{\rm end}=-H_{\rm end}^{2}$ but does not determine
$H_{\rm end}$ itself.

If the accumulated duration is dominated by the preceding large-$y$
quasi-de Sitter stage, with the exit interval contributing only
subleading corrections, equation~~\eqref{N_approx_z} implies
$N_{\rm tot}\simeq y_{\rm end}$ at leading order. Combining this
relation with equation~\eqref{a-N}, we obtain

\begin{equation}\label{a-N-2}
\frac{|\delta a|}{a_{*}}
\simeq
\frac{|\alpha-1|h_{\rm end}}{N_{\rm tot}}.
\end{equation}

The same relations connect the Taylor-control parameter to the total
duration measured from the operational onset. Combining
equations~\eqref{exit_eta_bound} and~\eqref{a-N-2} gives
\begin{equation}\label{N_lower_eta}
N_{\rm tot}
\gtrsim
\frac{5|\alpha-1|h_{\rm end}}{2\eta}.
\end{equation}

To illustrate the Taylor-control criterion, consider a 5\% tolerance for 
the quadratic correction, $\eta=0.05$. Equation~\eqref{exit_eta_bound} then gives
\begin{equation}\label{delta-num}
    \frac{|\delta a|}{a_{*}}\lesssim 2\times10^{-2}.
\end{equation}

Within this leading-order matching approximation,
equations~\eqref{a-N-2}--\eqref{delta-num}  are background-level consistency relations.
They show that, once the operational onset and the endpoint are
specified, the Taylor-control condition constrains the relation among
$\alpha$, $\eta$, $N_{\rm tot}$, and the dimensionless exit rate
$h_{\rm end}$. These relations neither provide a direct observational
constraint on $\alpha$ nor determine $N_{\rm tot}$ independently of
the chosen onset normalization, endpoint, and exit dynamics.


\subsection{Full-background evolution and local contractivity of the
radiation-like entrance}

\label{full-radiation-stability}

The existence of the exact large-$a$ solution~\eqref{a-rad-real} does not, by
itself, establish that the regular trajectory emerging from inflation
reaches that branch, determine its response to neighboring homogeneous
background trajectories, or quantify how long the subsequent evolution
remains close to the radiation-like behavior. We therefore complement
the asymptotic analysis of Subsection~\ref{p_equal_4} in three steps. 

First, we integrate the background equation with the complete
$p=4$ interpolating force, without using the small-$a$ or
large-$a$ approximation. 
Second, we study linear homogeneous perturbations across the
entrance domain and test their finite-time contraction in
relative variables. Third, we analyze the growing and decreasing relative
modes about the large-$a$ radiation-like branch. The latter analysis
provides a matching-dependent semi-analytic estimate for the duration
$\Delta N_{\rm rad}$ and a corresponding constraint on
$|\alpha-1|$. This is a homogeneous background analysis and should not
be confused with an analysis of scalar or tensor cosmological
perturbations.

Neglecting the ordinary pressureless matter contribution under the dominance
condition~\eqref{rad_dom_con}, substituting the complete force~\eqref{K-def} into the background
equation~\eqref{acc-eq}, using the variables $\zeta=a/a_c$ introduced above equation~\eqref{G_y}
and $y=\sqrt{\mathcal F_0}\,t$ introduced in equation~\eqref{a_exact_bessel}, we obtain
\begin{equation}
 \zeta''(y)
 +\left(\frac{\alpha-1}{y}\right)\zeta'(y)
 =
 f(\zeta),
 \qquad
 f(\zeta)
 \equiv
 \zeta\left[\frac{1-\zeta^4}{(1+\zeta^4)^2}\right].
 \label{full-p4-dimensionless}
\end{equation}
Here and throughout this subsection, a prime denotes differentiation with
respect to $y$, with $\mathcal F_0>0$. 

Let $\zeta_b(y)$ denote the background trajectory obtained by continuing the
regular expanding branch through the end of inflation. A neighboring
homogeneous trajectory can be written as
\begin{equation} \label{zeta-perturbation}
 \zeta(y)
 =
 \zeta_b(y)+\delta\zeta(y),
 \qquad
 \left|\frac{\delta\zeta}{\zeta_b}\right|\ll1.
\end{equation}
Substituting equation~\eqref{zeta-perturbation} into
equation~\eqref{full-p4-dimensionless} and retaining only terms linear in
$\delta\zeta$ gives
\begin{equation} \label{linear-zeta-perturbation}
 \delta\zeta''
 +\left(\frac{\alpha-1}{y}\right)\delta\zeta'
 -f_{,\zeta}(\zeta_b(y))\delta\zeta
 =
 0,
\end{equation}

where 
\begin{equation}
f_{,\zeta}(\zeta)\equiv\frac{df}{d\zeta}
 =
 \frac{1-12\zeta^4+3\zeta^8}
 {(1+\zeta^4)^3}.
 \label{force-gradient}
\end{equation}

Since the derivative is evaluated along the time-dependent background
trajectory $\zeta_b(y)$, the quantity
$f_{,\zeta}\!\left(\zeta_b(y)\right)$ is generally a function of $y$.
Equation~\eqref{linear-zeta-perturbation} is therefore a linear differential equation with
time-dependent coefficients. The sign of
$f_{,\zeta}\!\left(\zeta_b(y)\right)$ determines whether the complete
force provides a local restoring contribution. The boundaries of the
restoring interval follow from $f_{,\zeta}(\zeta)=0$, giving
\begin{equation}
 \zeta_{\pm}
=
 \left(
 \frac{6\pm\sqrt{33}}{3}
 \right)^{1/4},
 \qquad
 \zeta_-\simeq0.5401,
 \qquad
 \zeta_+\simeq1.4066.
 \label{restoring-boundaries}
\end{equation}
Consequently, along the background trajectory,
$f_{,\zeta}\bigl(\zeta_b(y)\bigr)<0$ whenever $\zeta_-<\zeta_b(y)<\zeta_+$. Defining $\omega_{\rm tr}^2(y)\equiv
-f_{,\zeta}(\zeta_b(y))>0$ inside this interval,
the perturbation equation takes the time-dependent oscillator form
\begin{equation} \label{transition-oscillator}
 \delta\zeta''
 +\left(\frac{\alpha-1}{y}\right)\delta\zeta'
 +\omega_{\rm tr}^2(y)\delta\zeta
 =
 0.
\end{equation}

Since $y>0$, the coefficient of the first-derivative term in
equation~\eqref{transition-oscillator},
$c_{\rm fr}(y)=(\alpha-1)/y$, is positive when $\alpha>1$; in this
case the fractional kernel supplies positive damping to the homogeneous
background perturbation. Independently, the radiation-like
solution~\eqref{a-rad-real} has a real amplitude only when $\alpha<3/2$.
Combining these requirements gives the parameter interval
\begin{equation}\label{sufficient-local-alpha}
1<\alpha<\frac{3}{2}.
\end{equation}

Equation~\eqref{sufficient-local-alpha} is a sufficient local condition ensuring, during the
passage through the restoring domain, a restoring force, positive
fractional damping, and a real radiation-like amplitude. It should not
be interpreted as a necessary condition or as a statement of global
asymptotic stability. The finite-time transfer calculation below
provides the stronger numerical test and shows contraction of relative
perturbations also for representative values below unity.
For a numerically stable treatment over many e-folds, we introduce the
logarithmic variables
\begin{equation} \label{logarithmic-variables}
 n\equiv\ln\zeta,
 \qquad
 s\equiv\ln y,
 \qquad
 v\equiv\frac{dn}{ds}
       =\frac{y\zeta'}{\zeta}
       =tH.
\end{equation}
Using the definitions for $x$ and $G$ given respectively by equations~\eqref{K-def} and \eqref{G_y}, 
$x=\zeta^4= e^{4n}$ and $ G(n)\equiv {(1-x)}/{(1+x)^2}$,
equation~\eqref{full-p4-dimensionless} becomes the first-order, nonautonomous system
\begin{equation}
 \label{log-background-system}
 \frac{dn}{ds}
 =
 v,
\qquad
 \frac{dv}{ds}
 =
 y^2G(n)-[v^2+(\alpha-2)v].
    \end{equation}

For the numerical propagation of the same homogeneous perturbation
introduced in equation~\eqref{zeta-perturbation}, we express it in the logarithmic
phase-space variables. At linear order,
\begin{equation}
\delta n=\frac{\delta\zeta}{\zeta_b},
\qquad
\delta v=\frac{d(\delta n)}{ds}
=\frac{y}{\zeta_b}\delta\zeta'-v_b\delta n.
\end{equation}
Equivalently, we write $n=n_b+\delta n$ and $v=v_b+\delta v$.
Linearizing equation~\eqref{log-background-system} then gives the variational system
\begin{equation} \label{linear-log-system}
 \frac{d}{ds}
 \begin{pmatrix}
  \delta n\\[1mm]
  \delta v
 \end{pmatrix}
 =
 \mathsf A(s)
 \begin{pmatrix}
  \delta n\\[1mm]
  \delta v
 \end{pmatrix},
\end{equation}
where the time-dependent Jacobian matrix is
\begin{equation}
 \mathsf A(s)
 =
 \begin{pmatrix}
  0
  &
  1
  \\[1mm]
  y^2G_{,n}(n_b)
  &
  -\bigl(2v_b+\alpha-2\bigr)
 \end{pmatrix},
 \qquad
 G_{,n}
 =
 \frac{4x(x-3)}{(1+x)^3},
 \label{variational-matrix}
\end{equation}
where all background quantities in $\mathsf A(s)$ are evaluated along
$(n_b(s),v_b(s))$.

Let $s_-$ and $s_+$ denote the logarithmic times at which the
monotonically expanding background enters and leaves the restoring
interval, respectively: $\zeta_b(e^{s_-})=\zeta_-$ and $\zeta_b(e^{s_+})=\zeta_+$.

The fundamental matrix $\mathsf M(s,s_-)$ associated with the
variational system~\eqref{linear-log-system} is defined by
\begin{equation}
 \frac{d\mathsf M}{ds}
 =
 \mathsf A(s)\mathsf M,
 \qquad
 \mathsf M(s_-,s_-)=\mathsf I_2 .
 \label{fundamental-matrix}
\end{equation}
It propagates the unscaled perturbation vector according to
\begin{equation}\label{matrixA-n-v}
\begin{pmatrix}
\delta n(s)\\
\delta v(s)
\end{pmatrix}
=
\mathsf M(s,s_-)
\begin{pmatrix}
\delta n(s_-)\\
\delta v(s_-)
\end{pmatrix}.
\end{equation}
Thus, $\mathsf M(s_+,s_-)$ represents the finite-time propagation
of the unscaled perturbations across the restoring interval.
To compare dimensionless relative perturbations having the same physical
meaning at the two boundaries, we introduce
\begin{equation} \label{relative-transfer-vector}
 \mathbf X(s)
 \equiv
 \begin{pmatrix}
  \delta n\\[1mm]
  \delta v/v_b
 \end{pmatrix}
 =
 \begin{pmatrix}
  \delta a/a\\[1mm]
  \delta H/H
 \end{pmatrix},
 \qquad
 \mathbf X(s_+)
 =
 \mathsf M_{\rm tr}\mathbf X(s_-).
\end{equation}
The vector $\mathbf{X}$ expresses the same perturbations in
relative form. At fixed $s$, $\delta n=\delta a/a$ and,
since $v=tH$, $\delta v/v_b=\delta H/H$.
The transfer matrix in equation~~\eqref{relative-transfer-vector} is, therefore,
\begin{equation}\label{Mat-D}
 \mathsf M_{\rm tr}
 =
 \mathsf D_+
 \mathsf M(s_+,s_-)
 \mathsf D_-^{-1},
 \qquad
 \mathsf D_{\pm}
 =
 \operatorname{diag}
 \left(
 1,\frac{1}{v_b(s_{\pm})}
 \right),
\end{equation}
where the diagonal matrices
$\mathsf D_\pm$ only convert between the unscaled and relative
representations at the entrance and exit boundaries.

The singular values of $\mathsf M_{\rm tr}$ are defined as the
nonnegative square roots of the eigenvalues of
$\mathsf M_{\rm tr}^{\rm T}\mathsf M_{\rm tr}$: $\sigma_i(\mathsf M_{\rm tr})
=
\sqrt{
\lambda_i\!\left(
\mathsf M_{\rm tr}^{\rm T}\mathsf M_{\rm tr}
\right)}$.
Here, $\lambda_i\!\left(\mathsf M_{\rm tr}^{\rm T}\mathsf M_{\rm tr}\right)$,
with $i=1,2$, denotes the $i$th eigenvalue of the symmetric
positive-semidefinite matrix
$\mathsf M_{\rm tr}^{\rm T}\mathsf M_{\rm tr}$. Hence,
$\lambda_i\geq0$ and the singular values $\sigma_i$ are real and
nonnegative. The notation $\lambda_i$ used here refers only to a
matrix eigenvalue.

For a two-component relative perturbation vector
$\mathbf X=(X_1,X_2)^{\rm T}$, the Euclidean norm is defined by
\[
\|\mathbf X\|_2
=
\sqrt{X_1^2+X_2^2}
=
\sqrt{
\left(\frac{\delta a}{a}\right)^2+
\left(\frac{\delta H}{H}\right)^2
}.
\]

In particular, the largest singular value admits the equivalent
definition

\[
\sigma_{\max}(\mathsf M_{\rm tr})
=
\max_{\mathbf X\neq\mathbf 0}
\frac{\|\mathsf M_{\rm tr}\mathbf X\|_2}{\|\mathbf X\|_2}.
\]
It therefore measures the maximum possible finite-time amplification
of any dimensionless relative perturbation in the Euclidean norm.
Consequently,
\begin{equation}
 \sigma_{\max}(\mathsf M_{\rm tr})<1
 \quad\Longrightarrow\quad
 \|\mathbf X(s_+)\|_2
 <
 \|\mathbf X(s_-)\|_2
 \quad
 \text{for every nonzero linear perturbation}.
 \label{finite-time-contraction}
\end{equation}

This criterion establishes finite-time contraction across the
restoring interval in the chosen relative Euclidean norm; by itself,
it does not constitute a proof of global asymptotic stability.

We integrated equations~\eqref{log-background-system} and
\eqref{fundamental-matrix} along a regular background
trajectory with matching data chosen to give $N_{\rm tot}=60$
according to equation~\eqref{N-tot}. This value serves as an illustrative
numerical benchmark and is not identified with $N_{\rm CMB}$.
Table~1 reports the singular values of the transfer matrix and
the duration $\Delta N_{\rm rad}$ of the contiguous interval
satisfying $|w_{\rm eff}-1/3|<0.1$.

\begin{table}[ht]
\centering
\caption{Finite-time contraction of homogeneous relative perturbations
across the restoring interval
$\zeta_-<\zeta<\zeta_+$ for the complete $p=4$ background with
$N_{\rm tot}=60$.}
\label{full-background-transition}
\small
\begin{tabular}{c c c c}
\hline
$\alpha$ & $\sigma_{\max}$ & $\sigma_{\min}$ &
$\Delta N_{\rm rad}$\\
\hline
$0.999$ & $0.92500$ & $0.33931$ & $3.16$\\
$0.990$ & $0.92492$ & $0.33934$ & $2.00$\\
$0.900$ & $0.92414$ & $0.33967$ & $0.88$\\
\hline
$1.010$ & $0.92509$ & $0.33927$ & $1.89$\\
$1.050$ & $0.92544$ & $0.33912$ & $1.17$\\
$1.100$ & $0.92588$ & $0.33895$ & $0.89$\\
\hline
\end{tabular}
\end{table}

For example, for $\alpha=1.1$, the physical transfer matrix is
\begin{equation}
 \mathsf M_{\rm tr}
 \simeq
 \begin{pmatrix}
  0.5280 & 0.3762\\
  -0.7398 & 0.0673
 \end{pmatrix},
 \qquad
 \bigl(
 \sigma_{\max},
 \sigma_{\min}
 \bigr)
 =
 \bigl(
 0.9259,
 0.3389
 \bigr).
 \label{representative-transfer-matrix}
\end{equation}
The inequality~\eqref{finite-time-contraction} is therefore satisfied.
An independent scan over $ 0.1\leq\alpha<1$ and $1<\alpha\leq1.49$ gives
$0.9169
 \lesssim
 \sigma_{\max}
 \lesssim
 0.9294$, with no loss of finite-time contraction. Repeating the calculation over
$10\leq N_{\rm tot}\leq100$ leads to the same qualitative result.

As numerical consistency checks, the determinant of the unscaled fundamental
matrix agrees with Liouville's identity,
\begin{equation}
 \det\mathsf M
 =
 \exp
 \left[
  \int_{s_-}^{s_+}
  {\rm tr}\,\mathsf A(s)\,ds
 \right],
\end{equation}
to a relative accuracy better than $10^{-12}$. Moreover, finite differences
between two neighboring complete background solutions reproduce the
variational matrix to an accuracy better than $10^{-7}$.

In what follows, we estimate how long the regular trajectory remains close to the
radiation-like branch. This question is distinct from the finite-time
contraction across the entrance domain discussed above.

In the large-$\zeta$ regime, equation~\eqref{full-p4-dimensionless}
reduces to
\begin{equation}
 \zeta''+\frac{\alpha-1}{y}\zeta'
 \simeq -\frac{1}{\zeta^{3}},
 \label{large-zeta-reduced}
\end{equation}
whose radiation-like solution can be written as
\begin{equation}
 \zeta_{\rm r}(y)=A_{\rm r}y^{1/2},
 \qquad
 A_{\rm r}^{4}=\frac{4}{3-2\alpha}.
 \label{dimensionless-radiation-branch}
\end{equation}
Therefore, a positive real amplitude requires $\alpha<3/2$, in
agreement with the reality condition obtained in Subsection~\ref{p_equal_4}.

To examine departures from this branch, we write
\begin{equation}
 \zeta(y)=\zeta_{\rm r}(y)\,[1+\chi(y)],
 \qquad |\chi|\ll1 .
 \label{radiation-relative-perturbation}
\end{equation}
Substitution into equation~\eqref{large-zeta-reduced} and
linearization in $\chi$ gives
\begin{equation}
 \chi''+\frac{\alpha}{y}\chi'
 -\frac{3-2\alpha}{y^{2}}\chi=0 .
 \label{radiation-relative-equation}
\end{equation}
Accordingly, $\chi\propto y^{q_\pm}$, where
\begin{equation}
 q_{\pm}(\alpha)=
 \frac{1-\alpha\pm
 \sqrt{\alpha^{2}-10\alpha+13}}{2}.
 \label{radiation-mode-indices}
\end{equation}
Within the radiation-like reality domain $\alpha<3/2$, the two roots $q_{\pm}(\alpha)$
have opposite signs, so that the solution contains one decreasing mode
and one growing mode. Near, but not at, the special point $\alpha=1$,
one has $q_+\simeq1$ and $q_-\simeq-1$.

During the radiation-like interval, the regular trajectory remains
close to $\zeta_r\propto y^{1/2}$. Its duration is therefore
\begin{equation} \label{radiation-y-efold}
 \Delta N_{\rm rad}
\equiv
\ln\left[
\frac{\zeta_b(y_{\rm rad,out})}
     {\zeta_b(y_{\rm rad,in})}
\right]
\simeq
\frac12\ln\left(
\frac{y_{\rm rad,out}}{y_{\rm rad,in}}
\right),
\end{equation}
where $y_{\rm rad,in}$ and $y_{\rm rad,out}$ denote, respectively,
the first and last points of the contiguous interval satisfying
$|w_{\rm eff}-1/3|<0.1$. 

For the reference radiation-like branch, defining
\begin{equation}
 \Delta N(y)\equiv
 \ln\left[
 \frac{\zeta_{\rm r}(y)}
      {\zeta_{\rm r}(y_{\rm rad,in})}
 \right],
\end{equation}
the growing relative mode evolves from the entrance point according to
\begin{equation}\label{radiation-growing-mode}
 \frac{\chi_{\rm grow}(y)}
      {\chi_{\rm grow}(y_{\rm rad,in})}
 =
 \left(\frac{y}{y_{\rm rad,in}}\right)^{q_+}
 =
 \exp\left[2q_+\Delta N(y)\right].
 \end{equation}

The smooth numerical matching to the preceding regular trajectory
motivates, near but not at the special point $\alpha=1$, the
leading-order parametrization $\left|\chi_{\rm grow}(y_{\rm rad,in})\right|
\simeq B_{\rm in}\,|\alpha-1|$,
where $B_{\rm in}$ depends on the background matching data but varies
slowly near unity. Let $\chi_\star$ denote the relative-departure
threshold at which the trajectory leaves the operationally defined
radiation-like interval. Using equation~\eqref{radiation-growing-mode}, the exit condition is
therefore $B_{\rm in}|\alpha-1|\,
\exp\!\left(2q_+\Delta N_{\rm rad}\right)
\simeq\chi_\star $.
Defining the dimensionless matching coefficient
$\mathcal{C}_{\Delta N}\equiv\chi_\star/B_{\rm in}$ then gives the
semi-analytic estimate
\begin{equation}\label{radiation-duration-bound}
 \Delta N_{\rm rad}\simeq
 \frac{1}{2q_{+}(\alpha)}
 \ln\left(\frac{ \mathcal C_{_{\rm \Delta N}}}{|\alpha-1|}\right),
 \qquad
 |\alpha-1|\lesssim
 \mathcal C_{_{\rm \Delta N}}
 e^{-2q_{+}(\alpha)\Delta N_{\rm rad}} .
 \end{equation}

Here, $\mathcal{C}_{\Delta N}$ is fixed by the background matching data
and by the tolerance adopted to define the radiation-like interval.
Within the near-unity domain, this coefficient varies slowly and is
treated as constant at leading order. For the criterion
$|w_{\rm eff}-1/3|<0.1$ and the normalization used in Table~\ref{full-background-transition}, a
one-parameter fit to the numerical results gives
$\mathcal{C}_{\Delta N}\simeq0.54$. Since
$q_+(\alpha)\simeq1$ sufficiently close to unity, equation~\eqref{radiation-duration-bound}
reduces to
\begin{equation} \label{radiation-duration-near-unity}
 \Delta N_{\rm rad}\simeq
 \frac{1}{2}
 \ln\left(\frac{0.54}{|\alpha-1|}\right),
 \qquad
 |\alpha-1|\lesssim
 0.54\,e^{-2\Delta N_{\rm rad}} .
\end{equation}

For illustration, extrapolating equation~\eqref{radiation-duration-near-unity} within the adopted
matching prescription to $\Delta N_{\rm rad}\gtrsim25$ gives
\begin{equation}\label{alpha_cons_rad}
|\alpha-1|\lesssim
0.54e^{-50}
\simeq1.0\times10^{-22}.
\end{equation}
This relation is a matching-dependent semi-analytic estimate, rather
than a model-independent analytic or observational bound on $\alpha$.

The present background analysis does not treat $T_\alpha$ as an
independent fundamental fluid. As shown explicitly in \ref{AppendixA}, the
effective density and pressure are determined by the combined
fractional energy
$E_\alpha=T_\alpha+\Phi_\alpha$, whereas the separate additive
constants entering $T_\alpha$ and $\Phi_\alpha$ are fixed by the global
energy convention and by matching to the same trajectory. The
numerical evolution presented here is obtained directly from the
complete acceleration equation and therefore incorporates
simultaneously the dynamical effects of the time kernel and of the full
interpolating potential. No independent cancellation or resetting of
$T_\alpha$ is imposed at the end of inflation.

This estimate does not require
$\mathcal F_0(\alpha)\to0$ as $\alpha\to1$. As clarified below
equation~\eqref{K-def}, no such limiting behavior is imposed. Only in the exact
special case $\alpha=1$ is the fractional sector switched off.

We therefore distinguish three complementary results. First, the
complete interpolating force produces a smooth entrance into the
radiation-like regime. 

Second, the transfer calculation shows finite-time contraction
of relative homogeneous perturbations across the entrance domain
for the parameter values examined, including values below unity. Third, the duration of the subsequent radiation-like
interval is controlled by the growing relative mode and increases as
$\alpha$ approaches unity, according to the matching-dependent
estimate~\eqref{radiation-duration-bound}. All these conclusions are obtained within the
spatially flat sector, $\mathcal K=0$, and do not require the introduction of
an effective spatial-curvature contribution.



\section{Conclusions and Discussion}
\label{concl}

In this work, we have investigated the dynamics of the early universe
within the the emergent
$\Lambda$CDM cosmology developed in Ref.~\cite{rasouli2026emergent}.

The emergent $\Lambda$CDM cosmology is conceptually rooted in
generalized non-trivial measure constructions. In particular, it is
inspired by frameworks involving fractal geometries and scale-dependent
effective dimensions~\cite{calcagni2010fractal,calcagni2010quantum}, and therefore provides a non-trivial
extension of cosmological dynamics beyond the scope of standard
Newtonian descriptions. In this framework, both the time-dependent
fractional kernel, which gives rise to a fractional kinetic contribution,
and the fractional potential constitute the two fundamental ingredients
of the theory. As shown in Ref.~\cite{rasouli2026emergent}, their simultaneous presence is
essential for the internal consistency of the framework and for
recovering the corresponding relativistic cosmological background
dynamics. Moreover, the viable parameter space is significantly
constrained by the underlying structure, naturally favoring values of
the fractional parameter close to the standard Newtonian value,
$\alpha=1$, consistent with the constraints reported in
Refs.~\cite{rasouli2026emergent,rasouli2026evolution,Rasouli:2026zyq}.

Guided by the physical requirements, we identified a minimal
fractional potential for describing early-universe dynamics
within this framework. The potential alone, however, does not
determine the cosmological evolution. The effective dynamics
depend on the combined contribution of the time-dependent
fractional kernel and the fractional potential through the
total fractional energy $E_\alpha=T_\alpha+\Phi_\alpha$.
Here, $T_\alpha$ arises from the time-dependent kernel, while
$\Phi_\alpha$ denotes the fractional potential. Their relative
importance changes during cosmic evolution, giving rise to
different dominant dynamical regimes.

The corresponding fractional force, defined in equation~\eqref{K-phi},
is characterized by a small set of parameters with distinct
physical roles. In equation~\eqref{phi-alpha-def}, $a_c$ sets the characteristic
scale, while $p$ determines the shape and asymptotic behavior
of the force. Their relation to the force-zero scale $a_*$
is given by equation\eqref{x-star}. For $p=4$, which yields the
radiation-like power-law solution, this relation reduces
to $a_*=a_c$. In this case, $a_c$ is therefore the scale
at which the fractional force vanishes and changes sign.
In the exact special Newtonian case $\alpha=1$,
$\Phi_0(1)=\mathcal F_0(1)=0$, and the fractional sector is switched off.
This specialization does not impose the limiting behavior
$\mathcal F_0(\alpha)\to0$ as $\alpha\to1$.

We have shown that, for $x\equiv(a/a_c)^p\ll1$, the fractional
force is approximately constant and positive, giving rise to
quasi-de Sitter inflation through the background dynamics.
Within this constant-force approximation, the background equation
admits an exact solution, and requiring finite $a$, $\dot a$,
and $\ddot a$ at the initial boundary selects a unique regular
branch for $0<\alpha<2$, with $\alpha\neq1$.
The same trajectory evolves continuously from the regular
pre-inflationary regime toward quasi-de Sitter inflation.
These regimes thus describe the early-time and large-$y$
behavior of a single regular background solution.
An exact nonlinear analysis of homogeneous Hubble deviations
further establishes the asymptotic attractor character of
the inflationary regime within the admissible expanding sector
of the reduced constant-force dynamics.

Along this regular branch, $a(0)=a_{\min}>0$, $H(0)=0$,
and $\dot H(0)=\mathcal F_0/\alpha$ are finite. Although the fractional
coefficient $\gamma_\alpha=(\alpha-1)/t$ diverges as $t\to0^+$,
its contribution $\gamma_\alpha H$ remains finite because
$H\propto t$ near the initial boundary. The background equations
and the principal curvature invariants therefore remain finite
there, so that the corresponding spatially flat FLRW background
has no curvature singularity at this boundary.

For $p=4$, the reduced large-$a$ equation admits the radiation-like
solution $a(t)\propto t^{1/2}$ with $w_{\rm eff}=1/3$.
The complete-background analysis in Subsection~\ref{full-radiation-stability} shows that
the regular trajectory emerging from inflation enters a finite
radiation-like interval. Relative homogeneous perturbations
contract across the entrance domain for the parameter values
examined, while the growing relative mode controls the subsequent
duration. The radiation-like branch is therefore not a generic
late-time attractor. These results concern the background
evolution and do not provide a microscopic mechanism for
reheating or thermalization.

The end of accelerated expansion is defined by
$\ddot a(t_{\rm end})=0$. Expanding the fractional force around
$a_*$ gives a local estimate of the exit displacement
$\varepsilon_{\rm ex}=(a_{\rm end}-a_*)/a_*$, with its validity
controlled by the tolerance $\eta$. Subsection~\ref{smooth_exit} derives
same-epoch consistency relations within the quasi-de Sitter
regime and separate exit relations that retain the dependence
on $H_{\rm end}$. The quasi-de Sitter relations are not
extrapolated to the end of inflation. The total duration
$N_{\rm tot}$ depends on the operational onset, the endpoint,
and the initial and matching data; it is neither fixed by
$\alpha$ alone nor identified with the conventional CMB
interval of approximately 50--60 e-folds. These results are
background-level consistency conditions, rather than
independent predictions for $\alpha$ or $N_{\rm tot}$.

The cosmological horizon problem remains an open question in
this framework. Although the underlying fractional Newtonian
formulation does not intrinsically define a causal light-cone
structure, the background admits an effective FLRW representation
(see Subsection~\ref{nonsingularity}). In this representation, a particle horizon
can be defined, allowing the nominal and sufficient inflation
conditions to be examined along the lines of
Refs.~\cite{levin1995gravity,rasouli2014noncommutative,Rasouli:2016syh,rasouli2019kinetic}. Such an analysis has not been carried out
here and lies beyond the scope of the present study.

It is important to emphasize that the present work should be viewed as
the early-universe sector of a broader fractional cosmological program.
The primary objective is not merely to construct an inflationary
background, but to demonstrate that the same fractional Newtonian
framework previously introduced in Ref.~\cite{rasouli2026emergent}
can provide a unified effective description of the background
cosmological evolution, extending from a regular pre-inflationary stage
to the present epoch within a single
theoretical setting. Based on the general background equation~\eqref{acc-eq} and
the unified interpolating potential introduced in Section~5 of
Ref.~\cite{rasouli2026emergent}, together with the early-universe
potential~\eqref{phi-alpha-def} proposed in the present work (omitting the
radiation contribution of Ref.~\cite{rasouli2026emergent} to
avoid double counting), one may construct a single effective fractional
potential encompassing the successive background regimes of cosmic
evolution.

Cosmological perturbations in the fractional framework have previously
been investigated during the matter-dominated epoch using a fluid--flow
approach \cite{rasouli2026evolution}, yielding the observational bound
$\alpha<1.07$ from large-scale structure formation, consistent with the
parameter range considered here. Inflationary perturbations, however,
have not yet been developed within the present framework. Consequently,
references below to perturbative properties or observational predictions
of conventional inflationary models are included only for general
context and should not be interpreted as a perturbative comparison with
the fractional model. A dedicated analysis of primordial scalar and
tensor perturbations will be required to connect the inflationary
background obtained here quantitatively with standard inflationary
observables.

The inflationary background emerging in the fractional framework shares
several features with conventional inflationary scenarios, including
standard scalar-field models
\cite{linde1983chaotic,linde2000inflationary,brandenberger2010cosmology}
and their generalized versions
\cite{rasouli2018inflationary,rasouli2022noncommutativity,rasouli2022inflation,shah2026inflationary},
Starobinsky inflation
\cite{galante2015unity,starobinsky1980new,vilenkin1985classical},
and the class of $\alpha$-attractor\footnote{We emphasize
that the fractional parameter $\alpha$ employed throughout the present
work is conceptually unrelated to the parameter conventionally denoted
by $\alpha$ in $\alpha$-attractor models.}
models
\cite{kallosh2013universality,kallosh2013superconformal,iacconi2023novel,del2025primordial}. The comparison below is therefore intended to clarify the physical and
conceptual position of the fractional framework relative to these
conventional inflationary scenarios, while remaining strictly at the
level of background dynamics.

In standard single-field inflationary scenarios, the accelerated
expansion of the early universe is driven by a fundamental scalar
degree of freedom $\phi$ evolving under a potential $V(\phi)$.
Within this framework, the scalar potential sets the inflationary
energy scale, controls the slow-roll dynamics, and determines the
mechanism responsible for the exit from inflation. The number of
$e$-folds, the amplitude of primordial fluctuations, and the spectral
properties of perturbations are governed by the structure of
$V(\phi)$. From a conceptual perspective, inflation is attributed to
an additional physical degree of freedom, namely the inflaton field,
whose dynamics dominate the early universe.

Starobinsky inflation originates from a modification of the gravitational
action through the inclusion of an $R^2$ term, where $R$ is the Ricci
scalar. Although it is formulated purely geometrically in the Jordan
frame, it can be recast in the Einstein frame as an effective scalar-field
theory with a specific potential. In this case, a characteristic mass
scale associated with the higher-curvature correction sets the
inflationary scale and controls the quasi-de Sitter phase. The exit from
inflation arises dynamically as the system departs from the slow-roll
regime described in the effective scalar representation.

Similarly, $\alpha$-attractor models are formulated in terms of
scalar-field dynamics, often embedded in superconformal or supergravity
constructions. In these models, the parameter $\alpha$ characterizes
the curvature scale of the scalar-field target space and governs the
approach to universal attractor predictions. Inflation is again driven
by an explicit scalar degree of freedom, with $\alpha$ playing a central
role in determining the inflationary dynamics and the resulting
observational predictions.

In contrast, the fractional cosmological framework developed here does
not invoke any fundamental or effective scalar field as the source of
inflation. Instead, accelerated expansion emerges dynamically from the
combined contribution of the time-dependent fractional kernel and the
fractional potential. The fractional parameter $\alpha$ does not
represent a new propagating degree of freedom; rather, it quantifies the
structural departure from the special Newtonian
case. Inflation arises as a
regime of the modified gravitational dynamics itself.
The end of inflation is determined by the dynamical balance
between the fractional force and the time-dependent
friction/anti-friction term, with a locally controlled
displacement from the force-zero scale.
The quasi-de Sitter analysis relates the accumulated e-folds,
the fractional parameter $\alpha$, and the effective equation
of state at the same epoch. Separately, the local exit analysis
relates the exit displacement to $\alpha$ and $H_{\rm end}$.
These consistency relations do not independently determine
the total duration of inflation.

Together, these results reveal a fundamental conceptual distinction.
While scalar-field, Starobinsky, and $\alpha$-attractor scenarios
attribute inflation to an explicit scalar degree of freedom
(fundamental or geometrically induced), the fractional model interprets
inflation as an emergent gravitational phenomenon arising from a
minimal deformation of Newtonian dynamics.

\section{Acknowledgments}
The author sincerely thanks the Reviewers for their careful reading of
the manuscript and for their valuable comments and suggestions, which
helped improve the clarity and quality of the work.
The author acknowledges
 the FCT grant \textbf{UID/212/2025} Centro de Matem\'{a}tica 
e Aplica\c{c}\~{o}es da Universidade da Beira Interior plus
the COST Actions CA23130 (Bridging high and low energies in search of
quantum gravity (BridgeQG)) and CA23115 (Relativistic Quantum Information (RQI)).

\appendix
\section{Fractional energy contributions and normalizations}
\label{AppendixA}

In this appendix, we evaluate the fractional kinetic contribution,
the fractional potential, the combined fractional energy, and the
conserved total mechanical energy along the regular solution.
We clarify how the additive integration constants are fixed
and how the adopted energy normalization is related to the
spatial-curvature parameter $\mathcal{K}$.

Here, the exact regular solution refers to the branch derived
in Subsection~\ref{regular_branch} under the approximations
$\mathcal{F}(a)\simeq F_0$ and $\rho_m\simeq0$,
without imposing the quasi-de Sitter condition
$|\dot H|\ll H^2$. Its exactness is restricted to this reduced
constant-force equation.

\subsection{Exact regular trajectory}
\label{ER-traj}

The regular solution of the reduced background equation~\eqref{transition_a_equation}
is given by equation \eqref{a_pre_exact}. For notational convenience, we define
\begin{eqnarray}\label{A-B-notation}
 \mathcal{A}_{\alpha}(t)
\equiv
{}_0F_1\left(
;\frac{\alpha}{2};
\frac{\mathcal{F}_{0}t^2}{4}
\right),
\qquad
\mathcal{B}_{\alpha}(t)
\equiv
{}_0F_1\left(
;1+\frac{\alpha}{2};
\frac{\mathcal{F}_{0}t^2}{4}
\right).
\end{eqnarray}
Using equation~\eqref{A-B-notation}, the regular solution~\eqref{a_pre_exact} and its first derivative can then be written as
\begin{eqnarray}\label{app-areg-compact}
a_{\rm reg}(t)
=
a_{\min}\mathcal{A}_{\alpha}(t),
\qquad
\dot a_{\rm reg}(t)
=
\left(\frac{a_{\min}\,\mathcal{F}_{0}}{\alpha}\right)\,
t\,\mathcal{B}_{\alpha}(t).
\end{eqnarray}
Consequently, the standard kinetic energy contribution evaluated along the
regular trajectory is
\begin{align}\label{app-Treg}
T^{(\rm reg)}(t)
\equiv
\frac{1}{2}\dot a_{\rm reg}^{\,2}(t)
=
\left(\frac{a_{\min}^{2}\mathcal{F}_{0}^{\,2}}
     {2\alpha^2}\right)
t^2\,\mathcal{B}_{\alpha}^{\,2}(t).
\end{align}

Along the regular branch, the defining relation for the fractional
kinetic contribution gives
$\dot T_\alpha^{(\mathrm{reg})}
=\gamma_\alpha\dot a_{\mathrm{reg}}^{\,2}$.
A useful exact expression for $T_{\alpha}^{(\rm reg)}$ can be obtained
without explicitly integrating the hypergeometric functions. From
equation~\eqref{transition_a_equation}, one finds
\begin{align}
\frac{d}{dt}
\left[
\frac{\mathcal{F}_{0}}{2}a_{\rm reg}^{\,2}
-\frac{1}{2}\dot a_{\rm reg}^{\,2}
\right]
=
\mathcal{F}_{0}a_{\rm reg}\dot a_{\rm reg}
-\dot a_{\rm reg}\ddot a_{\rm reg}
=
\gamma_\alpha(t)\dot a_{\rm reg}^{\,2}.
\label{app-Ta-identity}
\end{align}
Integrating this identity and using $a_{\mathrm{reg}}(0)=a_{\min}$ and
$\dot a_{\mathrm{reg}}(0)=0$ gives
\begin{equation}\label{app-Ta-reg}
T_{\alpha}^{(\rm reg)}(t)
=
C_T^{(\rm reg)}(\alpha)
+
\frac{\mathcal{F}_{0}}{2}
\left[
a_{\rm reg}^{\,2}(t)-a_{\min}^{2}
\right]
-
\frac{1}{2}\dot a_{\rm reg}^{\,2}(t),
\end{equation}
where $C_T^{(\rm reg)}(\alpha) \equiv T_{\alpha}^{(\rm reg)}(0)$.
Equivalently, in terms of the hypergeometric functions~\eqref{A-B-notation},  equation~\eqref{app-Ta-reg} can be rewritten as
\begin{align}\label{app-Ta-reg-explicit}
T_{\alpha}^{(\rm reg)}(t)
={}
C_T^{(\rm reg)}(\alpha)
+
\frac{\mathcal{F}_{0}a_{\min}^{2}}{2}
\left[
\mathcal{A}_{\alpha}^{\,2}(t)-1
\right]
-
\frac{a_{\min}^{2}\mathcal{F}_{0}^{\,2}t^2}
     {2\alpha^2}
\mathcal{B}_{\alpha}^{\,2}(t).
\end{align}

Within the same reduced constant-force approximation, the fractional
potential \eqref{phi-alpha-def}, evaluated along the regular branch, takes the form
\begin{equation}
\Phi_{\alpha}^{(\rm reg)}(t)
=
-\frac{\mathcal{F}_{0}}{2}
a_{\rm reg}^{\,2}(t)
+
C_\Phi^{(\rm reg)}(\alpha),
\label{app-Phi-reg}
\end{equation}
where $C_\Phi^{(\rm reg)}(\alpha)$ denotes the additive potential
constant in this representation. This constant does not affect the
force or the reduced background equation, but it contributes to the
fractional energy.

Writing
$C_{\mathrm{reg}}(\alpha)\equiv
C_T^{(\mathrm{reg})}(\alpha)+C_\Phi^{(\mathrm{reg})}(\alpha)$, the fractional energy evaluated along the exact regular branch becomes

\begin{align}\label{app-Ea-reg}
E_{\alpha}^{(\rm reg)}(t)
\equiv
T_{\alpha}^{(\rm reg)}(t)
+
\Phi_{\alpha}^{(\rm reg)}(t)
=
C_{\rm reg}(\alpha)
-\frac{\mathcal{F}_{0}a_{\min}^{2}}{2}
-\frac{1}{2}\dot a_{\rm reg}^{\,2}(t).
\end{align}

The cancellation of the terms proportional to
$a_{\rm reg}^{2}(t)$ in equation~\eqref{app-Ea-reg} is a direct
consequence of the reduced equation of motion and is independent of
the particular representation of the regular solution.

Using equations~\eqref{app-Treg} and
\eqref{app-Ea-reg},  the conserved
effective energy $\mathcal{E}^{(\rm reg)}
\equiv
T^{(\rm reg)}(t)
+
E_{\alpha}^{(\rm reg)}(t)$ is
\begin{equation}
\mathcal{E}^{(\rm reg)}
=
C_{\rm reg}(\alpha)
-
\frac{\mathcal{F}_{0}a_{\min}^{2}}{2}.
\label{app-Etot-reg}
\end{equation}
This expression is independent of time, as required by the conserved
energy relation.

According to the curvature convention adopted in the main text, $\mathcal K=-2\mathcal{E}^{(\rm reg)}$.
It follows that
\begin{equation}
\mathcal K
=
\mathcal{F}_{0}a_{\min}^{2}
-
2C_{\rm reg}(\alpha).
\label{app-K-Creg}
\end{equation}
Therefore, restriction to the spatially flat sector, $\mathcal K=0$, fixes the
sum of the two additive constants according to
\begin{equation}
C_{\rm reg}(\alpha)
=
C_T^{(\rm reg)}(\alpha)
+
C_\Phi^{(\rm reg)}(\alpha)
=
\frac{\mathcal{F}_{0}a_{\min}^{2}}{2}
.
\label{app-flat-condition}
\end{equation}

Thus, spatial flatness does not require the two additive constants to
vanish separately. Rather, it fixes only their sum.

The potential normalization adopted in the main text,
$C_\Phi^{(\mathrm{reg})}(\alpha)=0$, therefore fixes
\begin{equation}
C_T^{(\mathrm{reg})}(\alpha)
=
T_\alpha^{(\mathrm{reg})}(0)
=
\frac{\mathcal F_0a_{\min}^2}{2},
\end{equation}
whereas the alternative convention
$C_T^{(\rm reg)}(\alpha)=0$ would require $C_\Phi^{(\rm reg)}(\alpha)
=
{\mathcal{F}_{0}a_{\min}^{2}}/{2}$.
These choices are related by a redistribution of the additive
constant between the fractional kinetic and potential contributions;
the conserved total energy depends only on their sum.

After imposing equation~\eqref{app-flat-condition}, the conserved
energy and the fractional energy simplify to
\begin{equation}
\mathcal{E}^{(\rm reg)}=0,
\qquad
E_{\alpha}^{(\rm reg)}(t)
=
-\frac{1}{2}\dot a_{\rm reg}^{\,2}(t)
=
-T^{(\rm reg)}(t)
.
\label{app-flat-energies}
\end{equation}

For completeness, before imposing $\mathcal K=0$, the effective fractional
density can be written as
\begin{align}
\rho_{\alpha}^{(\rm reg)}(t)
&=
-\frac{3E_{\alpha}^{(\rm reg)}(t)}
       {4\pi G a_{\rm reg}^{2}(t)}
=
\frac{3H_{\rm reg}^{2}(t)}{8\pi G}
-
\frac{3\mathcal{E}^{(\rm reg)}}
     {4\pi G a_{\rm reg}^{2}(t)}
\nonumber\\\nonumber\\
&=
\frac{3H_{\rm reg}^{2}(t)}{8\pi G}
+
\frac{3\mathcal K}{8\pi G a_{\rm reg}^{2}(t)}.
\label{app-rho-reg}
\end{align}
The second term has the usual $a^{-2}$ curvature-like scaling. In the
spatially flat sector it vanishes identically after the additive
constants are fixed through
equation~\eqref{app-flat-condition}.

\subsection{Pre-inflationary regime}

The pre-inflationary expressions follow by expanding the exact
regular trajectory as $t\to0^+$. The superscript ``pre'' denotes
this early-time expansion, rather than a separate solution.

Using equation \eqref{a_pre_expansion}, we obtain the standard kinetic energy contribution associated with the pre-inflationary regime as 
\begin{align}\label{app-Tpre}
T^{(\rm pre)}(t)
={}\frac{a_{\min}^{2}\mathcal{F}_{0}^{\,2}}
     {2\alpha^2}t^2
+
\frac{a_{\min}^{2}\mathcal{F}_{0}^{\,3}}
     {2\alpha^2(\alpha+2)}t^4
+\mathcal{O}(t^6).
\end{align}

We write
$C_T^{(\mathrm{pre})}(\alpha)=C_T^{(\mathrm{reg})}(\alpha)$,
$C_\Phi^{(\mathrm{pre})}(\alpha)=C_\Phi^{(\mathrm{reg})}(\alpha)$,
and $C_{\mathrm{pre}}(\alpha)=C_{\mathrm{reg}}(\alpha)$.
These are the same integration constants as in the regular
solution.

Expanding equation~\eqref{app-Ta-reg}, the fractional kinetic
contribution becomes
\begin{align}
T_{\alpha}^{(\rm pre)}(t)
={}&
C_T^{(\rm pre)}(\alpha)
+
\frac{(\alpha-1)a_{\min}^{2}\mathcal{F}_{0}^{\,2}}
     {2\alpha^2}t^2
\nonumber\\
&+
\frac{(\alpha-1)a_{\min}^{2}\mathcal{F}_{0}^{\,3}}
     {4\alpha^2(\alpha+2)}t^4
+\mathcal{O}(t^6).
\label{app-Ta-pre}
\end{align}

Similarly, the fractional potential is
\begin{align}
\Phi_{\alpha}^{(\rm pre)}(t)
={}&
C_\Phi^{(\rm pre)}(\alpha)
-\frac{\mathcal{F}_{0}a_{\min}^{2}}{2}
-\frac{a_{\min}^{2}\mathcal{F}_{0}^{\,2}}
     {2\alpha}t^2
\nonumber\\
&-
\frac{a_{\min}^{2}\mathcal{F}_{0}^{\,3}(\alpha+1)}
     {4\alpha^2(\alpha+2)}t^4
+\mathcal{O}(t^6).
\label{app-Phi-pre}
\end{align}

Adding equations~\eqref{app-Ta-pre} and
\eqref{app-Phi-pre}, we obtain
\begin{align}
E_{\alpha}^{(\rm pre)}(t)
={}&
C_{\rm pre}(\alpha)
-\frac{\mathcal{F}_{0}a_{\min}^{2}}{2}
-\frac{a_{\min}^{2}\mathcal{F}_{0}^{\,2}}
     {2\alpha^2}t^2
\nonumber\\
&-
\frac{a_{\min}^{2}\mathcal{F}_{0}^{\,3}}
     {2\alpha^2(\alpha+2)}t^4
+\mathcal{O}(t^6).
\label{app-Ea-pre}
\end{align}
The total pre-inflationary energy is therefore
\begin{align}
\mathcal{E}^{(\rm pre)}
&\equiv
T^{(\rm pre)}(t)
+
E_{\alpha}^{(\rm pre)}(t)
\nonumber\\
&=
C_{\rm pre}(\alpha)
-\frac{\mathcal{F}_{0}a_{\min}^{2}}{2}
=
-\frac{\mathcal K}{2}.
\label{app-Etot-pre}
\end{align}
The time-dependent terms cancel through the retained order,
consistently with conservation of the effective mechanical energy.

At the initial boundary,
\[
T^{(\mathrm{pre})}(0)=0,\qquad
T_\alpha^{(\mathrm{pre})}(0)=C_T^{(\mathrm{pre})}(\alpha),
\qquad
\Phi_\alpha^{(\mathrm{pre})}(0)
=C_\Phi^{(\mathrm{pre})}(\alpha)-\frac{\mathcal F_0a_{\min}^2}{2}.
\]
With the spatially flat normalization~\eqref{app-flat-condition}, these values give
\[
E_\alpha^{(\mathrm{pre})}(0)=\mathcal{E}^{(\mathrm{pre})}=0,
\qquad
T_\alpha^{(\mathrm{pre})}(0)
+\Phi_\alpha^{(\mathrm{pre})}(0)=0.
\]
Thus, spatial flatness fixes the sum of the additive constants.
Setting both constants to zero would violate this normalization
when $\mathcal F_0a_{\min}^2\neq0$.

\subsection{Matching to the quasi-de Sitter representation}
\label{app:regular-inflation-matching}

The constants appearing in the regular and inflationary
representations are not independent. Using the large-$y$
expansion of the regular-branch Hubble parameter given
in equation~\eqref{H-assymp}, we obtain
\begin{align}
\frac{1}{2}\dot a_{\rm reg}^{\,2}(t)
&=
\frac{a_{\rm reg}^{2}(t)}{2}H_{\rm reg}^{2}(t)
\nonumber\\
&\simeq
\frac{a_{\rm reg}^{2}(t)}{2}
\left[
\mathcal{F}_{0}
+
\frac{\sqrt{\mathcal{F}_{0}}(1-\alpha)}{t}
\right],
\label{app-Treg-large}
\end{align}
where terms of order $t^{-2}$ and smaller in the expansion
of $H_{\rm reg}^2$ have been omitted.

Using equation~\eqref{app-Ea-reg}, the large-time asymptotic
form of the regular-branch fractional energy is therefore
\begin{align}
E_{\alpha}^{(\rm reg)}(t)
\simeq{}&
\mathcal{E}^{(\rm reg)}
\nonumber\\
&-
\frac{a_{\rm reg}^{2}(t)}{2}
\left[
\mathcal{F}_{0}
+
\frac{\sqrt{\mathcal{F}_{0}}(1-\alpha)}{t}
\right].
\label{app-Ereg-large}
\end{align}
Comparison with the inflationary expression in the main text,
together with the exact conserved-energy relation from~\ref{ER-traj}, identifies

\begin{equation}\label{app-Cinf-matching}
C_{\rm inf}(\alpha)
=
\mathcal{E}^{(\rm reg)}
=
C_{\rm reg}(\alpha)
-\frac{\mathcal{F}_{0}a_{\min}^{2}}{2}
=
-\frac{\mathcal K}{2}.
\end{equation}

Since both representations use the same global fractional
potential, their additive constants satisfy
$C_\Phi^{(\mathrm{reg})}(\alpha)=C_\Phi(\alpha)$.
Consequently,
\begin{equation}
C_T^{(\rm inf)}(\alpha)
=
C_T^{(\rm reg)}(\alpha)
-\frac{\mathcal{F}_{0}a_{\min}^{2}}{2}.
\label{app-CT-matching}
\end{equation}
Thus, $C_T^{(\rm inf)}$ is not an additional independent integration
constant. It is the constant appearing when the same regular
trajectory is represented in its large-time quasi-de Sitter form.

In the spatially flat sector considered here,
equation~\eqref{app-Cinf-matching} gives $C_{\mathrm{inf}}(\alpha)=0$.

Accordingly, in equations~\eqref{rho_inf} and~\eqref{p_inf}, the terms proportional to
$C_{\rm inf}(\alpha)/a_{\rm inf}^{2}$ in the inflationary density
and pressure vanish after the regular branch is consistently matched
to the spatially flat inflationary regime. Before imposing
$\mathcal K=0$, the same terms have the expected $a^{-2}$ curvature-like
scaling.

\subsection{Radiation-like branch and its energy normalization}
\label{app:radiation-energy}

We finally investigate the fractional-energy quantities associated with the
radiation-like solution of the reduced large-$a$ equation. For $p=4$, the behavior of the
fractional force is given by \eqref{K-expan-rad}. Neglecting the ordinary-matter
contribution in this regime, the corresponding radiation-like 
solution is represented by equation~\eqref{a-rad-real}. The following expressions are evaluated on this reference branch;
departures of the complete trajectory from it are studied
in Subsection~\ref{full-radiation-stability}.

The standard kinetic contribution on this branch is
$T^{(\mathrm{rad})}(t)=a_r^2/(8t)$.
The defining relation
$\dot T_\alpha^{(\mathrm{rad})}
=2\gamma_\alpha T^{(\mathrm{rad})}
=(\alpha-1)a_r^2/(4t^2)$
then gives
\begin{equation}
 T_{\alpha}^{(\rm rad)}(t)
 =
 C_{T}^{(\rm rad)}(\alpha)
 +
 \frac{(1-\alpha)a_{r}^{2}}{4t}.
 \label{app-rad-Talpha}
\end{equation}
Here, $C_{T}^{(\rm rad)}(\alpha)$ denotes the additive constant appearing
when the globally defined fractional kinetic contribution is represented
in its large-$a$ asymptotic form. It is not an additional independent
constant and must be fixed by matching to the preceding evolution.

In the same approximation, the fractional potential is
$\Phi_\alpha^{(\mathrm{rad})}
=-\mathcal F_0a_c^4/(2a_{\mathrm{rad}}^2)
+C_\Phi^{(\mathrm{rad})}(\alpha)$.
Using equation~\eqref{a-rad-real}, this becomes
\begin{equation}
 \Phi_{\alpha}^{(\rm rad)}(t)
 =
 C_{\Phi}^{(\rm rad)}(\alpha)
 -
 \frac{(3-2\alpha)a_{r}^{2}}{8t}.
 \label{app-rad-Phi}
\end{equation}
The superscript ``rad'' labels the large-$a$ representation; the additive
potential constant is inherited from the same fractional potential and
does not represent a new force contribution.

Defining $ C_{\rm rad}(\alpha)
 \equiv
 C_{T}^{(\rm rad)}(\alpha)
 +
 C_{\Phi}^{(\rm rad)}(\alpha)$, the fractional energy on the radiation-like branch becomes
\begin{align}
 E_{\alpha}^{(\rm rad)}(t)
 &\equiv
 T_{\alpha}^{(\rm rad)}(t)
 +
 \Phi_{\alpha}^{(\rm rad)}(t)
 \nonumber\\
 &=
 C_{\rm rad}(\alpha)
 -
 \frac{a_{r}^{2}}{8t}
 =
 C_{\rm rad}(\alpha)-T^{(\rm rad)}(t).
 \label{app-rad-Ealpha}
\end{align}
The dependence on $\alpha$ in the two time-dependent fractional
contributions combines precisely so that their sum is independent of
$\alpha$ apart from the normalization $a_{r}$.

The conserved total mechanical energy evaluated on this branch is
therefore
\begin{equation}
 {\cal E}^{(\rm rad)}
 \equiv
 T^{(\rm rad)}(t)+E_{\alpha}^{(\rm rad)}(t)
 =
 C_{\rm rad}(\alpha).
 \label{app-rad-total-energy}
\end{equation}
Using \eqref{rho-eff}, we obtain $ \mathcal K=-2C_{\rm rad}(\alpha)$. Consequently, assuming the effective spatially flat FLRW metric fixes
\begin{equation}
 C_{\rm rad}(\alpha)
 =
 C_{T}^{(\rm rad)}(\alpha)
 +
 C_{\Phi}^{(\rm rad)}(\alpha)
 =0.
 \label{app-rad-flat}
\end{equation}
Hence, on the spatially flat radiation-like branch,
\begin{equation}
 {\cal E}^{(\rm rad)}=0,
 \qquad
 E_{\alpha}^{(\rm rad)}(t)
 =
 -T^{(\rm rad)}(t)
 =
 -\frac{a_{r}^{2}}{8t}.
 \label{app-rad-flat-energy}
\end{equation}

Before imposing spatial flatness, equations \eqref{rho-eff} and \eqref{p-eff}
give the fractional energy density and pressure as
\begin{align} \label{app-rad-rho}
 \rho_{\alpha}^{(\rm rad)}(t)
 &=
 \frac{3}{32\pi Gt^{2}}
 -
 \frac{3C_{\rm rad}(\alpha)}
      {4\pi G a_{\rm rad}^{2}(t)}
=
 \frac{3}{32\pi Gt^{2}}
 +
 \frac{3\mathcal K}{8\pi G a_{\rm rad}^{2}(t)},
\end{align}
\begin{align}
 p_{\alpha}^{(\rm rad)}(t)
=
 \frac{1}{32\pi Gt^{2}}
 +
 \frac{C_{\rm rad}(\alpha)}
      {4\pi G a_{\rm rad}^{2}(t)}
=
 \frac{1}{32\pi Gt^{2}}
 -
 \frac{\mathcal K}{8\pi G a_{\rm rad}^{2}(t)}.
 \label{app-rad-pressure}
\end{align}
It follows that
\begin{equation}
 p_{\alpha}^{(\rm rad)}
 -
 \frac{1}{3}\rho_{\alpha}^{(\rm rad)}
 =
 \frac{C_{\rm rad}(\alpha)}
      {2\pi G a_{\rm rad}^{2}}
 =
 -\frac{\mathcal K}{4\pi G a_{\rm rad}^{2}}.
 \label{app-rad-EoS-general}
\end{equation}
Therefore, for the spatially flat case, we obtain
\begin{equation}
 \rho_{\alpha}^{(\rm rad)}(t)
 =
 \frac{3}{32\pi Gt^{2}},
 \qquad
 p_{\alpha}^{(\rm rad)}(t)
 =
 \frac{1}{32\pi Gt^{2}},
 \qquad
 w_{\alpha}^{(\rm rad)}
 =
 \frac{1}{3}.
 \label{app-rad-EoS-flat}
\end{equation}

For completeness, if the large-$a$ representation is introduced at a
matching time $t_{\rm m}$, equation~\eqref{app-rad-Talpha} gives
\begin{equation}
 C_{T}^{(\rm rad)}(\alpha)
 =
 T_{\alpha}(t_{\rm m})
 -
 \frac{(1-\alpha)a_{r}^{2}}{4t_{\rm m}}.
 \label{app-rad-matching}
\end{equation}

Thus, $C_T^{(\mathrm{rad})}$ is fixed by the preceding kinetic
history. Condition~\eqref{app-rad-flat} must therefore hold for matching
to the spatially flat radiation-like branch; it cannot be
enforced by independently readjusting the constants after
the evolution has been specified.

Equations~\eqref{app-rad-flat} and~\eqref{app-rad-EoS-flat}
complete the energy-normalization analysis in the post-inflationary
large-$a$ regime. Together with the regular-branch condition~\eqref{app-flat-condition},
they show that the additive constants in each asymptotic representation
are fixed by the conserved-energy convention and by matching to the
corresponding cosmological trajectory.

\section{Summary of time and e-fold conventions}
\label{AppendixB}

For clarity, we summarize the time, Hubble-flow, EoS parameter,
and e-fold conventions used in the analysis of inflation and
its dynamical exit. The exact regular solution of the reduced
constant-force equation defines the operational onset $t_i$,
whereas the endpoint $t_{\rm end}$ is determined from the full
background dynamics.

For comparison, an illustrative CMB horizon-crossing epoch
$t_{\rm CMB}$ may be included schematically as
\[
t_i
\longrightarrow t_{\rm CMB}
\xrightarrow{\;N_{\rm CMB}\;} t_{\rm end}
\longrightarrow \text{radiation-like regime}.
\]
Along the full background trajectory, the e-fold intervals satisfy
\[
N_{\rm tot}
=
\ln\left[\frac{a(t_{\rm CMB})}{a_{\rm on}}\right]
+N_{\rm CMB},
\qquad
N_{\rm CMB}\equiv
\ln\left[\frac{a_{\rm end}}{a(t_{\rm CMB})}\right].
\]
Here, $t_{\rm CMB}$ is introduced only to distinguish the CMB
interval from the total duration; no value of either
$t_{\rm CMB}$ or $N_{\rm CMB}$ is assumed in this work.\\


\begin{table}[htbp]
\begin{center}
\centering
\caption{Relevant epochs and scale factors}
\label{tab:Relevant epochs and scale factors}
\small
\begin{tabular}{|p{0.17\textwidth}|p{0.70\textwidth}|}
\hline
\textbf{Symbol} & \textbf{Definition and interpretation} \\
\hline

$a_*$
&
Force-zero scale defined by $\mathcal{F}(a_*)=0$. For $p=4$, one has
$a_*=a_c$. This scale is unrelated to the CMB horizon-crossing
notation.
\\
\hline

$t_i$
&
Operational onset time of the quasi-de Sitter stage, defined on the
exact regular branch of the reduced constant-force dynamics by
$\left|\epsilon_H^{(\mathrm{reg})}(t_i)\right|=0.1$. Equivalently,
$\left|\epsilon_H^{(\mathrm{reg})}(y_i;\alpha)\right|=0.1$, where
$y_i=\sqrt{\mathcal{F}_0}\,t_i$.
\\
\hline

$a_{\rm on}$
&
Scale factor at the operational onset:
$a_{\rm on}\equiv a_{\rm reg}(t_i)$. It fixes the onset normalization
used in defining the accumulated number of e-folds.
\\
\hline

$t_{\rm CMB}$
&
Horizon-crossing time of the selected CMB pivot scale. This epoch is
relevant only when the background evolution is connected to
inflationary observables. No numerical value of $t_{\rm CMB}$ is
assumed here.
\\
\hline

$t_{\rm end}$
&
Exact end of accelerated expansion, defined from the full background
dynamics by $\ddot a(t_{\rm end})=0$, or equivalently by
$\epsilon_H(t_{\rm end})=1$ and
$w_{\rm eff}(t_{\rm end})=-1/3$. It is not determined from the reduced
regular solution $a_{\rm reg}(t)$.
\\
\hline

$a_{\rm end}$
&
Scale factor at the exact endpoint:
$a_{\rm end}\equiv a(t_{\rm end})$. In general, $a_{\rm end}$ does not
coincide exactly with the force-zero scale $a_*$.
\\
\hline

\end{tabular}
\end{center}
\end{table}

\bigskip
\newpage

\begin{table}[htbp]
\begin{center}
\centering
\caption{Hubble-flow, EoS parameter, and exit quantities}
\label{tab:Hubble-flow, EoS parameter, and exit quantities}
\small
\begin{tabular}{|p{0.17\textwidth}|p{0.70\textwidth}|}
\hline
\textbf{Symbol} & \textbf{Definition and interpretation} \\
\hline

$\delta a$
&
Displacement of the exact endpoint from the force-zero scale:
$a_{\rm end}=a_*+\delta a$. For $p=4$, the relative exit displacement
is $\varepsilon_{\rm ex}\equiv\delta a/a_*$.
\\
\hline

$\eta$
&
Taylor-control parameter specifying the allowed magnitude of the
quadratic correction relative to the retained linear contribution:
$\frac{5}{2}\left|\delta a/a_*\right|\leq\eta$, with
$0<\eta\ll1$.
\\
\hline

$\epsilon_H^{(\mathrm{reg})}$
&
Hubble-flow parameter evaluated on the exact regular solution of the
reduced constant-force equation:
$\epsilon_H^{(\mathrm{reg})}\equiv
-\dot H_{\rm reg}/H_{\rm reg}^{2}$.
It is used to define $t_i$, but not $t_{\rm end}$.
\\
\hline

$\epsilon_H$
&
Hubble-flow parameter of the full background trajectory:
$\epsilon_H\equiv-\dot H/H^2$. The condition
$\epsilon_H(t_{\rm end})=1$ defines the exact end of accelerated
expansion.
\\
\hline

$w_{\rm eff}^{(\mathrm{inf})}(t)$
&
Effective EoS parameter of the inflationary background,
evaluated within the quasi-de Sitter regime. Its large-$y$ asymptotic
form should not be extrapolated to the exact endpoint.
\\
\hline

$w_{\rm eff}(t_{\rm end})$
&
Effective EoS parameter evaluated on the full
background trajectory at the exact endpoint:
$w_{\rm eff}(t_{\rm end})=-1/3$. It is distinct from
$w_{\rm eff}^{(\mathrm{inf})}(t)$.
\\
\hline

$y$
&
Dimensionless time variable:
$y\equiv\sqrt{\mathcal{F}_0}\,t$, with
$y_i=\sqrt{\mathcal{F}_0}\,t_i$ and
$y_{\rm end}=\sqrt{\mathcal{F}_0}\,t_{\rm end}$.
\\
\hline

$H_{\rm end}$
&
Hubble parameter evaluated on the full background trajectory at the
endpoint: $H_{\rm end}\equiv H(t_{\rm end})$. The endpoint condition
implies $\dot H_{\rm end}=-H_{\rm end}^{2}$ but does not determine
$H_{\rm end}$ itself.
\\
\hline

$h_{\rm end}$
&
Dimensionless endpoint Hubble parameter:
$h_{\rm end}\equiv H_{\rm end}/\sqrt{\mathcal{F}_0}$. No particular
numerical value of $h_{\rm end}$ is assumed.
\\
\hline

\end{tabular}
\end{center}
\end{table}

\bigskip
\newpage
\begin{table}[htbp]
\centering
\caption{E-fold conventions}
\label{tab:efold-conventions}
\renewcommand{\arraystretch}{1.25}
\begin{tabular}{|p{0.18\textwidth}|p{0.74\textwidth}|}
\hline
\textbf{Symbol} & \textbf{Definition and interpretation} \\
\hline

$N_{\rm on}(t)$
&
Number of e-folds accumulated along the exact regular branch
of the reduced constant-force equation from the operational
onset $t_i$ to time $t$:
\[
N_{\rm on}(t)\equiv
\int_{t_i}^{t}H_{\rm reg}(t')\,dt'
=
\ln\left[\frac{a_{\rm reg}(t)}{a_{\rm on}}\right],
\qquad
a_{\rm on}=a_{\rm reg}(t_i).
\]
\\
\hline

$N_{\rm CMB}$
&
Number of e-folds along the full background trajectory from
the selected CMB pivot-crossing epoch to the end of inflation:
\[
N_{\rm CMB}\equiv
\int_{t_{\rm CMB}}^{t_{\rm end}}H(t')\,dt'
=
\ln\left[\frac{a_{\rm end}}{a(t_{\rm CMB})}\right].
\]
No fixed numerical value of $N_{\rm CMB}$ is imposed.
\\
\hline

$N_{\rm tot}$
&
Total number of e-folds along the full background trajectory
from the operational onset to the end of inflation:
\[
N_{\rm tot}\equiv
\int_{t_i}^{t_{\rm end}}H(t')\,dt'
=
\ln\left(\frac{a_{\rm end}}{a_{\rm on}}\right).
\]
With the onset normalization $a(t_i)=a_{\rm on}$,
it satisfies
\[
N_{\rm tot}
=
\ln\left[\frac{a(t_{\rm CMB})}{a_{\rm on}}\right]
+N_{\rm CMB}.
\]
Its value depends on the initial and matching data and is
not identified with $N_{\rm CMB}$.
\\
\hline
\end{tabular}
\end{table}
\renewcommand{\arraystretch}{1.0}

\bibliographystyle{elsarticle-num}
\FloatBarrier
\bibliography{Frac-New-Inflation-Ref.bib} 
\end{document}